\documentclass[12pt, onecolumn, draftcls, a4paper]{IEEEtran}

\usepackage{epsfig}
\usepackage{subfigure}
\usepackage{amsmath}
\usepackage{float}
\usepackage{graphicx}
\usepackage{latexsym}
\usepackage{amssymb}
\usepackage{amsfonts}
\usepackage{cite}
\usepackage{url}
\usepackage{mathtools}
\usepackage{algorithm}
\usepackage{algpseudocode}

\usepackage{tikz}
\usepackage{pgfplots}
\usepackage{verbatim}
\usetikzlibrary{arrows}
\usetikzlibrary{positioning}
\usetikzlibrary{plotmarks}
\usetikzlibrary{decorations}
\usetikzlibrary{shapes}
\usetikzlibrary{plotmarks}

\usepackage{etex}
\usepackage[ngerman]{babel}
\usepackage[T1]{fontenc}
\usepackage{pgfplots}
\usepackage{flushend}
\usepackage{algorithm}
\usepackage{algpseudocode}

\DeclareMathOperator{\bb}{\mathbf{b}}

\DeclareMathOperator{\e}{\mathbf{e}}
\DeclareMathOperator{\f}{\mathbf{f}}
\DeclareMathOperator{\g}{\mathbf{g}}
\DeclareMathOperator{\h}{\mathbf{h}}
\DeclareMathOperator{\n}{\mathbf{n}}
\DeclareMathOperator{\p}{\mathbf{p}}

\DeclareMathOperator{\x}{\mathbf{x}}
\DeclareMathOperator{\y}{\mathbf{y}}
\DeclareMathOperator{\z}{\mathbf{z}}
\DeclareMathOperator{\A}{\mathbf{A}}
\DeclareMathOperator{\B}{\mathbf{B}}

\DeclareMathOperator{\cC}{\mathcal{C}}
\DeclareMathOperator{\bbC}{\mathbb{C}}
\DeclareMathOperator{\D}{\mathbf{D}}
\DeclareMathOperator{\cD}{\mathcal{D}}

\DeclareMathOperator{\E}{\mathbf{E}}
\DeclareMathOperator{\F}{\mathbf{F}}

\DeclareMathOperator{\G}{\mathbf{G}}
\DeclareMathOperator{\bH}{\mathbf{H}}
\DeclareMathOperator{\I}{\mathbf{I}}

\DeclareMathOperator{\nnull}{\mathcal{N}}

\DeclareMathOperator{\bP}{\mathbf{P}}
\DeclareMathOperator{\Q}{\mathbf{Q}}
\DeclareMathOperator{\cQ}{\mathcal{Q}}
\DeclareMathOperator{\R}{\mathbf{R}}

\DeclareMathOperator{\T}{\mathbf{T}}
\DeclareMathOperator{\V}{\mathbf{V}}

\DeclareMathOperator{\U}{\mathbf{U}}
\DeclareMathOperator{\W}{\mathbf{W}}
\DeclareMathOperator{\X}{\mathbf{X}}
\DeclareMathOperator{\Y}{\mathbf{Y}}
\DeclareMathOperator{\Z}{\mathbf{Z}}

\DeclareMathOperator{\0}{\mathbf{0}}

\DeclareMathOperator{\tr}{tr}

\DeclareMathOperator{\bvec}{vec}
\DeclareMathOperator{\diag}{diag}
\DeclareMathOperator{\rank}{rank}

\DeclareMathOperator{\IN}{{\scriptstyle IN}}

\DeclareMathOperator{\her}{{\scriptstyle H}}
\DeclareMathOperator{\tran}{{\scriptstyle T}}
\DeclareMathOperator{\mn}{{\scriptstyle \mathit{MN}}}
\DeclareMathOperator{\st}{{\scriptstyle such\; that}}
\DeclareMathOperator{\dB}{dB}

\DeclareMathOperator{\vGamma}{\boldsymbol \Gamma}

\newcommand{\boxeqn}[1]{
\[
\fbox{
\addtolength{\linewidth}{-2\fboxsep}%
\addtolength{\linewidth}{-2\fboxrule}%
\begin{minipage}{\linewidth}
\begin{equation*}
#1
\end{equation*}
\end{minipage}
}
\]
}

\newcommand{\PropBox}[1]{
\[
\fbox{
\addtolength{\linewidth}{-2\fboxsep}%
\addtolength{\linewidth}{-2\fboxrule}%
\begin{minipage}{\linewidth}
\begin{Proposition}
#1
\end{Proposition}
\end{minipage}
}
\]
}

\newtheorem{Remark}{Remark}

\newtheorem{Corollary}{Corollary}

\newtheorem{Proposition}{Proposition}

\title{Information Leakage Neutralization for the Multi-Antenna Non-Regenerative Relay-Assisted Multi-Carrier Interference Channel}
\author{Zuleita Ho, Eduard Jorswieck and Sabrina Gerbracht\\
Technische Universit\"{a}t Dresden\\
\{zuleita.ho, eduard.jorswieck, sabrina.gerbracht\}@tu-dresden.de
 \thanks{This work has been performed in the framework of the European research
project SAPHYRE, which is partly funded by the European Union under 
its FP7 ICT Objective 1.1 - The Network of the Future.}
\thanks{This work is supported by the German Research Foundation (DFG)
in the Collaborative Research Center 912 ``Highly Adaptive
Energy-Efficient Computing''.}
}

\begin{document}
 \maketitle
\begin{abstract}
In heterogeneous dense networks where spectrum is shared, users privacy remains one of the major challenges.
On a multi-antenna relay-assisted multi-carrier interference channel, each user shares the frequency and spatial resources with all other users.
 When the receivers are not only interested in their own signals but also in eavesdropping other users' signals, the cross talk on the frequency and spatial
 channels becomes information leakage. In this paper, we propose a novel secrecy rate enhancing relay strategy that utilizes both frequency and spatial resources,
termed as \emph{information leakage neutralization}. To this end,
the relay matrix is chosen such that 
the effective channel from the transmitter to the colluding eavesdropper is equal to the negative of
the effective channel over the relay to the colluding eavesdropper
and thus the information leakage to zero.
Interestingly, the optimal relay matrix in general is not block-diagonal which encourages users' encoding over the frequency channels.
We proposed two information leakage neutralization strategies, namely \emph{efficient information leakage neutralization} (EFFIN) 
and \emph{optimized information leakage neutralization} (OPTIN). EFFIN provides a simple and efficient design of relay processing matrix 
and precoding matrices at the transmitters in the scenario of limited power and computational resources. 
OPTIN, despite its higher complexity, provides a better sum secrecy rate performance
by optimizing the relay processing matrix and the precoding matrices jointly.
The proposed methods are shown to improve the sum 
secrecy rates over several state-of-the-art baseline methods.
\end{abstract}
\begin{keywords}
 Interference relay channel; Interference neutralization; Non-potent relay; Full-duplex relay; Amplify-and-forward relay; secrecy rate; worst-case secrecy rate;
frequency selective; multi-antenna systems; colluding eavesdroppers
\end{keywords}

\section{Introduction}
\begin{figure}\label{fig:ltecell}
 \begin{center}
 \resizebox{\linewidth}{4cm}{


\newcommand{\basestation}[2]{%
\begin{scope}[#1]
\draw (0,2) -- (1,6) -- (1,1.5);
\draw (2,2) -- (1,6) -- (1,1.5);
\draw (1,6) -- (1,6.4);
\draw (1,6.5) circle  (0.1cm);
\foreach \y in {1,2,3}{
\draw (0.8-0.1*\y,6.35-0.1*\y) .. controls (0.70-0.12*\y,6.5).. (0.8-0.1*\y,6.65+0.1*\y);
}
\foreach \y in {1,2,3}{
\draw (1.2+0.1*\y,6.35-0.1*\y) .. controls (1.3+0.12*\y,6.5).. (1.2+0.1*\y,6.65+0.1*\y);
}
\foreach \x in {0.1,0.6,1.1,1.6,2.1,2.6}{
\draw (0+0.26*\x,2+\x) -- (1,1.5+\x+.15*\x) -- (2-0.26*\x,2+\x) -- (1,2.5+\x-0.15*\x) -- cycle;
}
\draw (1,1.2) node {\tiny #2};
\end{scope}
}

\newcommand{\relay}[2]{%
\begin{scope}[#1]
\shadedraw[left color=white,right color=black!80] (0,0)--(0.5,3) -- (1.5,3) -- (2,0);
\filldraw[fill=black!30] (1,3) ellipse (0.5cm and .2cm);
\begin{scope}
\clip (-1,0) rectangle (3,-0.5);
\shadedraw[left color=white,right color=black!80] (1,0) ellipse (1cm and .4cm);
\end{scope}
\draw[thick] (1,3) -- (1,4);
\draw[thick] (0.3,4) -- (1.7,4);
\foreach \x in {0,1,2} {
\draw[thick] (0.3+0.7*\x,4) -- (0.3+0.7*\x,4.5);
\draw[thick] (0.3+0.7*\x,4.6) circle (0.1cm);
}
\draw (1,0.2) node {\tiny #2};
\end{scope}
}

\newcommand{\mobile}[2]{%
\begin{scope}[#1]
\begin{scope}
\clip (-0.05,-0.10) -- (-0.05,0.03)-- (1.3,0.03) -- (2,0.6) -- (2,-0.10) -- cycle;
\filldraw[fill=gray, draw=black, rounded corners=5pt] (0,-0.05) --(1.5,-0.05) -- (2.02,0.57) -- (0.5,0.55) -- cycle;
\end{scope}
\filldraw[fill=white, draw=black, rounded corners=5pt] (0,0) --(1.5,0) -- (2,0.6) -- (0.5,0.6) -- cycle;
\begin{scope}
\clip (0.44,1.65) -- (2.05,0.60) -- (2.05,1.65) -- cycle;
\filldraw[fill=gray, draw=black, rounded corners=5pt] (0.53,0.63) rectangle (2.03,1.63);
\end{scope}
\filldraw[fill=white, draw=black, rounded corners=5pt] (0.5,0.6) rectangle (2,1.6);
\filldraw[fill=black, draw=black] (0.65,0.75) rectangle (1.85,1.45);
\foreach \x in {1,2,3,4,5,6,7,8,9,10}{
\draw (0.25,0.1) -- (0.25+1.15/10*\x,0.1) -- (0.34+1.15/10*\x,0.2) -- (0.34,0.2) -- cycle;
\draw (0.4,0.3) -- (0.4+1.15/10*\x,0.3) -- (0.49+1.15/10*\x,0.4) -- (0.49,0.4) -- cycle;
}
\foreach \x in {1,2,3,4,5,6,7,8,9,10,11}{
\draw (0.28,0.2) -- (0.28+1.15/10*\x,0.2) -- (0.37+1.15/10*\x,0.3) -- (0.37,0.3) -- cycle;
\draw (0.45,0.4) -- (0.45+1.15/10*\x,0.4) -- (0.54+1.15/10*\x,0.5) -- (0.54,0.5) -- cycle;
}
\draw (0.8,-0.3) node {\tiny #2};
\end{scope}
}

\newcommand{\repeater}[2]{%
\begin{scope}[#1]
\filldraw[fill=white, draw=black] (0,0) rectangle (2,0.5);
\filldraw[fill=white, draw=black] (0,0.5) -- (1,1) -- (3,1) -- (2,0.5) -- cycle;
\filldraw[fill=white, draw=black]  (3,1) -- (3,0.5) -- (2,0) -- (2,0.5) -- cycle;
\draw (1.5,0.75) -- (1.5,1.4);
\draw (1.5,1.5) circle (0.1cm);
\foreach \x in {0,1,2,3,4,5}{
\draw (0.25+0.2*\x,0.35) circle (0.02cm);
}
\draw (1.8,0.25) circle (0.08);
\begin{scope}
\clip (1.79,0.25) -- (1.81,0.25) -- (1.81,0.18) -- (1.9,0.18) -- (1.9,0.4) -- (1.7,0.4) -- (1.7,0.18) -- (1.79,0.18) -- cycle;
\draw[thin] (1.8,0.25) circle (0.04);
\end{scope}
\draw[thin] (1.80,0.19) -- (1.80,0.24);
\foreach \y in {1,2,3}{
\draw (1.3-0.1*\y,1.35-0.1*\y) .. controls (1.2-0.12*\y,1.5).. (1.3-0.1*\y,1.65+0.1*\y);
}
\foreach \y in {1,2,3}{
\draw (1.7+0.1*\y,1.35-0.1*\y) .. controls (1.8+0.12*\y,1.5).. (1.7+0.1*\y,1.65+0.1*\y);
}
\draw (1.2,-0.3) node {\tiny #2};
\end{scope}
}

\begin{tikzpicture}[>=latex]

\begin{scope}
\filldraw[fill=orange!50!white, fill opacity=50] (0,-0.2) ellipse (5cm and 1cm);
\draw (-3,-0.3) node {\footnotesize LTE cell 1};
\filldraw[fill=orange!50!white, fill opacity=50] (7,0) ellipse (5cm and 1cm);
\draw (6,-0.5) node {\footnotesize LTE cell 2};
\filldraw[fill=orange!50!white, fill opacity=50] (2,1.3) ellipse (5cm and 1.1cm);
\draw (5.2,1.3) node {\footnotesize LTE cell 3};
\draw[gray] (0,-0.2) ellipse (5cm and 1cm);
\draw[gray] (7,0) ellipse (5cm and 1cm);
\draw[gray] (2,1.3) ellipse (5cm and 1.1cm);
\end{scope}

\relay{scale=0.4,shift={(7,1)}}{Relay};

\begin{scope}
\mobile{scale=0.5, shift={(-2.5,3.4)}}{UE 3};
\repeater{scale=0.4, shift={(0.8,5.1)}}{Repeater 3};
\basestation{scale=0.4, shift={(4.5,0.8)}}{BS 3};
\end{scope}

\begin{scope}
\mobile{scale=0.5, shift={(19cm,0.8cm)}}{UE 2};
\repeater{scale=0.4, shift={(20,-1.5)}}{Repeater 2};
\basestation{scale=0.4, shift={(17,-2)}}{BS 2};
\end{scope}

\begin{scope}
\mobile{scale=0.5, shift={(4,-1)}}{UE 1};
\repeater{scale=0.4, shift={(1,-2)}}{Repeater 1};
\basestation{scale=0.4, shift={(-1,-2.6)}}{BS 1};
\end{scope}


\draw[<-, thick ,red] (1.25,2.7) to [bend right=0] (1.85,2.9);
\draw[<-, thick ,red] (-0.15,2.35) to [bend left=0] (0.55,2.55);
\draw[<- , ultra thick, black!60] (-0.2,2.0) to [bend left=15] (2.8,2.15);
\draw[<-, thick, double distance=1pt] (2.9,2.3) to [bend right=15] (2.5,2.8);

\draw[<-, thick, red] (8.5,0.2) to [bend left=0] (7.5,1.7);
\draw[<-, thick, red] (9.5,0.35) to [bend right=0] (8.95,0.0);
\draw[<-, ultra thick, black!60] (9.7,1.2) to [bend right=17] (3.6,2.3);
\draw[<-, thick, double distance=1pt] (3.6,2.1) to [bend left=12] (6.8,1.8);

\draw[<-, thick, red] (0.8,0.05) to [bend left=0] (0.3,1.45);
\draw[<-, thick, red] (2.2,0.0) to [bend right=0] (1.35,-0.2);
\draw[<-, thick, double distance=1pt] (2.7,0.4) to [bend right=15] (3.25,1.95);
\draw[<-, thick, double distance=1pt] (2.9,2.0) to [bend left=15] (0.3,1.55);

\draw[<- ,dashed, thick, blue] (3.1,0.1) to [bend right=15] (6.8,1.7);

\end{tikzpicture}
}
\caption{Three overlapping LTE cells. The sum secrecy rates over the cells can be improved if a smart multi-antenna relay is introduced into the system. The 
emphasized arrows from BS 1 to the smart relay in the middle and then to UE 1 illustrate that desired signal strength (together
with the direct channel path in red) can be boosted
by choosing an appropriate relay strategy. The emphasized arrows from BS 2 to the smart relay and then to UE 1
illustrate that information leakage (shown by a dashed arrow in blue) can be neutralized by choosing the relay strategy appropriately.}
 \end{center}
\end{figure}
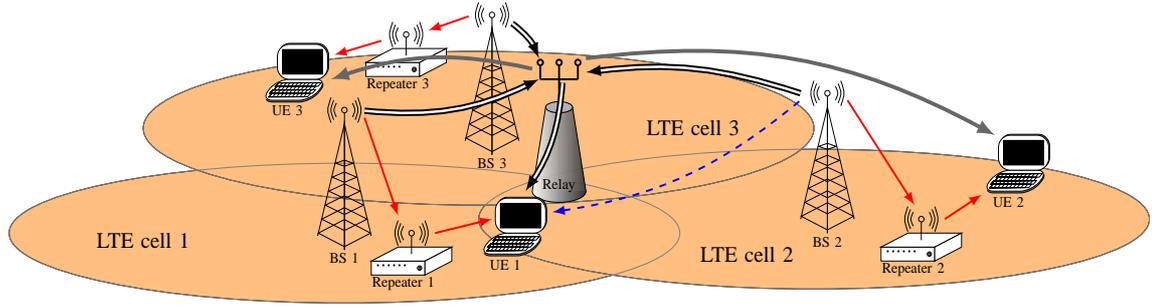

The trend of future wireless network systems is towards spectrum sharing over different wireless infrastructures such as
LTE networks, smart grid sensor networks and WiMAX networks. With isolated 
wireless infrastructures, such as multiple non-cooperating LTE cells (as shown in Figure \ref{fig:ltecell}), ensuring data security remains a major
technical challenge. While cryptography techniques are employed in most established communication standards,
physical layer security techniques provide an alternative approach when the communicating front-ends are of limited
computation capability and are not able to carry out standard cryptography methods such as symmetric key and asymmetric key encryption.
These applications include but are not limited to ubiquitous or pervasive computing \cite{Mattern2004}.

With the high demand of wireless applications in recent years, the issues of communication security become ever more important. 
Physical layer security techniques \cite{Liang2009a, Liu2009a, Bloch2011} provide an additional protection to the conventional 
secure transmission methods using cryptography. 
As early as four decades ago, the seminal work on the secrecy capacity on the wire-tap channel \cite{Wyner1975a} - the most fundamental model
consisting one source node, one destination node and one eavesdropper - started the era of research on physical layer security. 
Extensive analysis and designs have been conducted ever since; physical layer security results can be found in \cite{Liang2009a, Liu2009a, Bloch2011} 
and recent tutorial papers \cite{Shiu2011,Poor2012}.


With advantages such as increased cell coverage and transmission rates, relays are incorporated into the standards of current wireless 
infrastructures. The wireless resources in these systems are frequently shared by 
many users/subscribers and a potential malicious user in the system can lead to compromised confidentiality. Many novel 
strategies have been proposed to improve the secrecy in
\begin{itemize}
 \item  relay systems, including
cooperative jamming (CJ) \cite{Zheng2011,Dong2011,Huang2011}, noise-forwarding (NF) \cite{Lai2008}, a mixture of CJ and NF \cite{Bassily2012}, 
signal-forwarding strategies such as amplify-and-forward (AF) and decode-and-forward (DF) \cite{Petropulu2010,Ng2011,Bassily2012a}%
\footnote{All aforementioned works assume that the relays are cooperative and trusted. For secure transmission strategies with untrusted relays, 
please refer to \cite{He2010,Khodakarami2011,Jeong}.}. 
\item multi-carrier systems \cite{Jorswieck2008a, Wang2011b,Renna2012} 
and multi-carrier relay systems with external eavesdropper(s) \cite{Jeong2011, Ng2011}.
\end{itemize}
Yet, a joint optimization of secrecy rates over the frequency-spatial resources in a relay-assisted multi-user interference channel (with 
internal eavesdroppers) remains an open problem, as considered here.

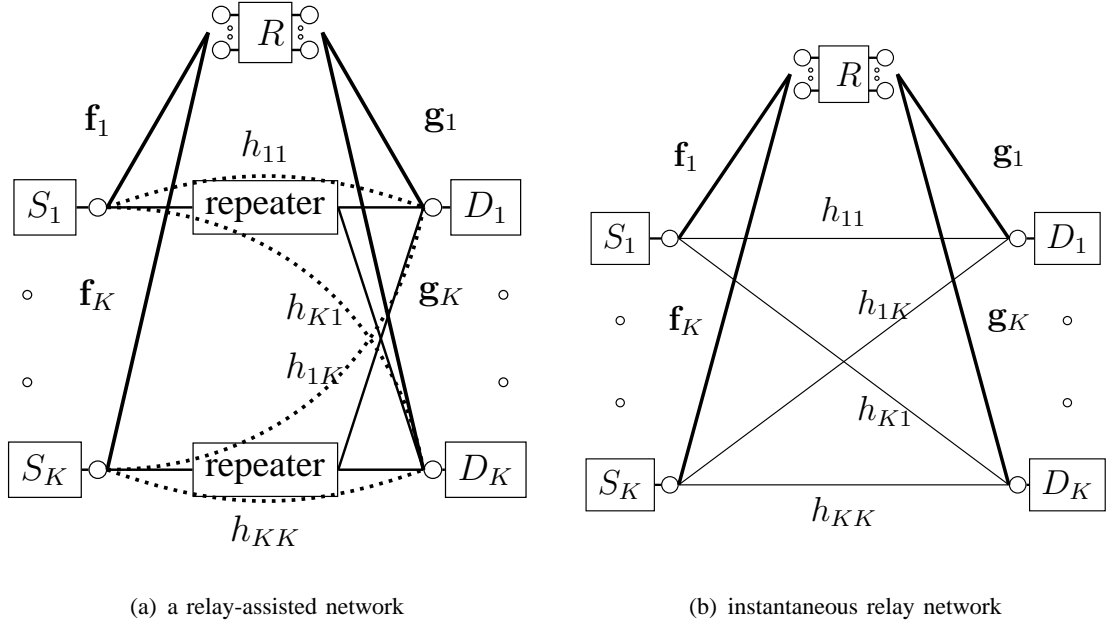
\begin{figure}
\begin{center}
\subfigure[a relay-assisted network]
{
\resizebox{0.45\linewidth}{!}{   
\begin{tikzpicture}
\node[rectangle,draw] (sk) at (0,0) {$S_K$};
\node[rectangle,draw] (s1) at (0,3) {$S_1$};
\node[circle,draw,inner sep=0, text width =0.1cm]  at (-0.2,2) {};
\node[circle,draw,inner sep=0, text width =0.1cm]  at (-0.2,1) {};

\node[circle,draw,inner sep=0, text width =0.2cm] (s1an) at (0.6,3) {};
\node[circle,draw,inner sep=0, text width =0.2cm] (skan) at (0.6,0) {};
\draw[thick] (s1) to (s1an);
\draw[thick] (sk) to (skan);

\node[rectangle,draw] (r2) at (2.5,0) {repeater};
\node[rectangle,draw] (r1) at (2.5,3) {repeater};

\draw[thick] (s1an.east) to (r1.west);
\draw[thick] (skan.east) to (r2.west);

\node[rectangle,draw] (dk) at (5,0) {$D_K$};
\node[rectangle,draw] (d1) at (5,3) {$D_1$};
\node[circle,draw,inner sep=0, text width =0.1cm]  at (5.2,2) {};
\node[circle,draw,inner sep=0, text width =0.1cm]  at (5.2,1) {};

\node[circle,draw,inner sep=0, text width =0.2cm] (d1an) at (4.4,3) {};
\node[circle,draw,inner sep=0, text width =0.2cm] (dkan) at (4.4,0) {};
\draw[thick] (d1) to (d1an);
\draw[thick] (dk) to (dkan);

\draw[thick] (r1.east) to (d1an.west);
\draw[thick] (r1.east) to (dkan.west);
\draw[thick] (r2.east) to (d1an.west);
\draw[thick] (r2.east) to (dkan.west);

\node[rectangle,draw, inner sep=0.2cm, text width =0.2cm] (r) at (2.5,5) {$R$};

\node[circle,draw,inner sep=0,text width =0.2cm] (ran1) at (2,5.2) {};
\node[circle,draw,inner sep=0,text width =0.2cm] (rank) at (2,4.8) {};
\draw[thick] (2.2, 5.2) -- (2.1, 5.2);
\draw[thick] (2.2, 4.8) -- (2.1, 4.8);
\node[circle,draw,inner sep=0,text width =0.05cm]  at (2.1, 5.05) {};
\node[circle,draw,inner sep=0,text width =0.05cm]  at (2.1, 4.95) {};
\node (r_an_l) at (2,5) {};

\node[circle,draw,inner sep=0,text width =0.2cm] (ran1) at (3,5.2) {};
\node[circle,draw,inner sep=0,text width =0.2cm] (rank) at (3,4.8) {};
\draw[thick] (2.8, 5.2) -- (2.9, 5.2);
\draw[thick] (2.8, 4.8) -- (2.9, 4.8);
\node[circle,draw,inner sep=0,text width =0.05cm]  at (2.9, 5.05) {};
\node[circle,draw,inner sep=0,text width =0.05cm]  at (2.9, 4.95) {};
\node (r_an_r) at (3,5) {};

\draw[very thick] (s1an.east) to (r_an_l.west);
\draw[very thick] (skan.east) to (r_an_l.west);
\draw[very thick] (r_an_r.east) to (d1an.west);
\draw[very thick] (r_an_r.east) to (dkan.west);

\node[very thick] at (4.5,2) {$\mathbf{g}_{K}$};
\node[very thick] at (0.6,2) {$\mathbf{f}_{K}$};
\node[very thick] at (4.5,4) {$\mathbf{g}_{1}$};
\node[very thick] at (0.6,4) {$\mathbf{f}_{1}$};

\path[very thick, dotted] (s1an.east) edge[bend left=20] node[anchor=south, above] {$h_{11}$} (d1an.west);
\path[very thick, dotted] (s1an.east) edge[bend left=40] node[anchor=north, below] {$h_{K1}$} (dkan.west);
\path[very thick, dotted] (skan.east) edge[bend right=20] node[anchor=north, below] {$h_{KK}$} (dkan.west);
\path[very thick, dotted] (skan.east) edge[bend right=40] node[anchor=south, above] {$h_{1K}$} (d1an.west);

\end{tikzpicture}
}
}
%
%
\subfigure[instantaneous relay network]
{
\resizebox{0.45\linewidth}{!}{
\begin{tikzpicture}

\node[rectangle,draw] (x1) at (-0.2,3) {$S_1$};
\node[circle,draw,inner sep=0, text width =0.1cm]  at (-0.2,2) {};
\node[circle,draw,inner sep=0, text width =0.1cm]  at (-0.2,1) {};
\node[rectangle,draw] (x2) at (-0.2,0) {$S_K$};

\node[circle,draw,inner sep=0, text width =0.2cm] (x1an) at (0.4,3) {};
\node[circle,draw,inner sep=0, text width =0.2cm] (x2an) at (0.4,0) {};
\draw[thick] (x1) to (x1an);
\draw[thick] (x2) to (x2an);

\node[rectangle,draw] (y1) at (5.2,3) {$D_1$};
\node[rectangle,draw] (y2) at (5.2,0) {$D_K$};
\node[circle,draw,inner sep=0, text width =0.1cm]  at (5.2,2) {};
\node[circle,draw,inner sep=0, text width =0.1cm]  at (5.2,1) {};

\node[circle,draw,inner sep=0, text width =0.2cm] (y1an) at (4.6,3) {};
\node[circle,draw,inner sep=0, text width =0.2cm] (y2an) at (4.6,0) {};
\draw[thick] (y1) to (y1an);
\draw[thick] (y2) to (y2an);

\draw (x1an.east) to node [anchor= south, above] {$h_{11}$} (y1an.west);
\draw (x1an.east) to  (y2an.west);
\draw (x2an.east) to node [anchor= north, below] {$h_{KK}$} (y2an.west);
\draw (x2an.east) to  (y1an.west);

\node at (3,2.2) {$h_{1K}$};
\node at (3,0.9) {$h_{K1}$};

\node[rectangle,draw, inner sep=0.2cm, text width =0.2cm] (r) at (2.5,5) {$R$};

\node[circle,draw,inner sep=0,text width =0.2cm] (ran1) at (2,5.2) {};
\node[circle,draw,inner sep=0,text width =0.2cm] (rank) at (2,4.8) {};
\draw[thick] (2.2, 5.2) -- (2.1, 5.2);
\draw[thick] (2.2, 4.8) -- (2.1, 4.8);
\node[circle,draw,inner sep=0,text width =0.05cm]  at (2.1, 5.05) {};
\node[circle,draw,inner sep=0,text width =0.05cm]  at (2.1, 4.95) {};
\node (r_an_l) at (2,5) {};

\node[circle,draw,inner sep=0,text width =0.2cm] (ran1) at (3,5.2) {};
\node[circle,draw,inner sep=0,text width =0.2cm] (rank) at (3,4.8) {};
\draw[thick] (2.8, 5.2) -- (2.9, 5.2);
\draw[thick] (2.8, 4.8) -- (2.9, 4.8);
\node[circle,draw,inner sep=0,text width =0.05cm]  at (2.9, 5.05) {};
\node[circle,draw,inner sep=0,text width =0.05cm]  at (2.9, 4.95) {};
\node (r_an_r) at (3,5) {};

\draw[very thick] (x1an.east) to (r_an_l.west);
\draw[very thick] (x2an.east) to (r_an_l.west);
\draw[very thick] (r_an_r.east) to (y1an.west);
\draw[very thick] (r_an_r.east) to (y2an.west);

\node[very thick] at (4.5,2) {$\mathbf{g}_{K}$};
\node[very thick] at (0.6,2) {$\mathbf{f}_{K}$};
\node[very thick] at (4.5,4) {$\mathbf{g}_{1}$};
\node[very thick] at (0.6,4) {$\mathbf{f}_{1}$};

\node[circle, inner sep=0, text width=0.2cm] at (0,-0.8) {};
\end{tikzpicture}
}
}
\caption{The wireless relay-assisted network with layer one repeaters and one smart relay is shown in 
subfigure (a). The dotted lines demonstrate the equivalent links between a source and a destination 
taking into account the presence of the repeaters. All paths from source to destination nodes take 
two time slots and links from source to relay and relay to destination take one time slot.  
The equivalent channel is established in subfigure (b) by replacing the relay as an instantaneous relay. 
Information going through the instantaneous relay arrives at the destinations  at the same time as over the direct links. \label{fig:irc}}
\end{center}
\end{figure}

We assume that the relay employs an amplify-and-forward (AF) strategy which provides flexibility in implementation as the relay is 
transparent to the modulation and coding schemes and induces negligible signal processing delays \cite{Berger2009}.
The novel notion of \emph{relay-without-delays}, also known as instantaneous relays if the relays are memoryless \cite{ElGamal2005,ElGamal2007,Cadambe2009a,Lee2011}, 
refers to relays that forward signals consisting of both current symbol and symbols in the past, instead of only the past symbols as in 
conventional relays. As shown in Figure \ref{fig:irc}, the instantaneous relay model provides a matching model of layer\nobreakdash-\hspace{0pt}1 repeaters connected networks (such 
as LTE networks) and helps us analyze the 
system performance of nowadays repeaters connected networks\footnote{In modern networks such as LTE, wireless links are often connected
 using boosters or layer-1 repeaters (simple amplifiers) 
\cite{Seidel2009}. If the time consumed for the signals to travel from a source to a repeater or from a repeater to a destination is counted as one unit, 
then the total time for the signal to travel from a source to a destination is two units - the same amount of time for the signal to travel from a source 
through a smart AF relay to a destination. }.


In order to provide secure transmission over relay-assisted multi-carrier networks, we propose a relay strategy termed as
\emph{information leakage neutralization} which by choosing relay forwarding strategies algebraically neutralizes 
information leakage from each transmitter in the network
to each eavesdropper on each frequency subcarrier. This method is adopted from a technique on relay networks, termed as
interference neutralization (IN).
IN has been applied to eliminate interference in various single-carrier systems, such as deterministic channels 
\cite{Mohajer2008, Mohajer2009a}, two-hop relay channels 
\cite{Berger2005, Rankov2007, Berger2009} and instantaneous relay channels \cite{Ho2011d}. Our prior work shows that IN 
is effective in improving secrecy rates in a two-hop wiretap channel \cite{Gerbracht2012}. The proposed method in this paper
differs from previous works above as the neutralization over multi-carrier systems is of high complexity. Another important difference is that
here the colluding eavesdroppers as well as the relays have multiple antennas. 

The contribution and outline of this manuscript are summarized as follows:
\begin{itemize}
 \item We transform a general and complicated sum secrecy rate optimization problem on a relay-assisted multi-carrier interference channel with 
mutually eavesdropping users to an optimization-ready formulation. Systematic optimization techniques can then be applied to 
solve for the sum-secrecy-rate-optimal relay strategies and precoding matrices at the transmitters.
\item An illustrative example is given in Section \ref{sec:example} for a basic setting to highlight the efficiency of information 
leakage neutralization.
 \item We propose a novel idea of information leakage neutralization strategies in Section \nolinebreak\ref{sec:in}. These strategies neutralize information 
leakage from each user to its colluding eavesdroppers on each frequency-spatial channel. The resulting secrecy rate expression is significantly simplified.
Detailed analyzes for the multi-carrier information leakage neutralization methods are provided. 
In particular, the minimum number
of antennas at the relay for complete information leakage neutralization is computed in Proposition 1. The required number of antennas depends on
the number of data streams sent by each user, the number of frequency subcarriers and the number of users in the system. Relevant to 
applications where relay power must be reserved, the minimum power at the relay required for information leakage neutralization is computed in Proposition 2.
\item We propose an efficient and simple information leakage neutralization strategy (EFFIN) which ensures
secure transmissions in the scenarios of limited power and computational resources at relay and transmitters. With sufficient power
at the relay, we propose an optimized information leakage neutralization technique (OPTIN) to maximize the secrecy rates while ensuring zero information
leakage.
\item The achievable secrecy rates from proposed strategies EFFIN and OPTIN are compared to several baseline strategies by numerical simulations
in Section \ref{sec:sim}. Baseline 1 
is a scenario where the relay is a layer-1 repeater and baseline 2 is a scenario with no relay. Simulation results show that the proposed 
strategies outperform the baseline strategies significantly in various operating SNRs. 
\end{itemize}

\subsection{Notations}
The set $\mathbb{C}^{a \times b}$ denotes a set of complex matrices of size $a$ by $b$ and is shortened to $\mathbb{C}^a$ when $a=b$. 
The notation $\mathcal{N}(\mathbf{A})$ is the null space of $\A$. 
The operator $\otimes$ denotes the Kronecker product. The superscripts $^{\tran}$, $^{\her}$, $^\dagger$ 
represent transpose, Hermitian transpose and Moore-Penrose inverse 
respectively whereas the superscript $^*$ denotes the conjugation operation. The Euclidean norm for scalars 
is written as $|.|$. The trace of matrix $\mathbf{A}$ is denoted as $\tr(\mathbf{A})$. 
Vectorization stacks the columns of a matrix $\A$ to form a long column vector denoted as $\bvec(\mathbf{A})$. 
The function $\mathcal{C}(\A)$ denotes the log-
determinant function of matrix $\A$, $\log \det \left(\A \right)$.
The identity and zero matrices of dimension $K\times K$ are written as $\I_K$ and $\0_
K$. The vector $\e_i$ represents a column vector with zero elements everywhere and one at the $i$-th position. 
The notation $[\A]_{ml}$ denotes the $m$-th row and $l$-th column element of the matrix $\A$. 
The notation $\p_{a:b}$, $0 \leq a\leq b\leq n$, denotes a vector which has elements $[p_a, p_{a+1},\ldots, p_b]$ where $\p=[p_1,\ldots,p_n]$.

\section{System Model}

In the following subsection, we give an example of a two-user interference relay channel in which the relay has two antennas and all nodes share two frequency
subcarriers. We shall illustrate that the conventional assumption of block diagonal relay matrix (which maximizes achievable rates in peaceful systems) cannot be
adopted a-priori when secrecy rates are considered.

\subsection{An example of two users on two frequencies with two antennas at the relay}\label{sec:example}
Transmitter $i$, $i=1,2$, transmits symbols $\x_i \in \bbC^{M \times 1}$ which are spread 
over $M$ frequency subcarriers by precoding matrix $\bP_i$. For the ease of notation, we assume that precoding matrix $\bP_i$
is a square matrix $\bP_i \in \bbC^{M}$. When user $i$ transmits $S_i\leq M$ symbols, then zeros are padded in $\x_i$ so that its
dimension is always $M \times 1$ and correspondingly zero columns are padded in $\bP_i$.
We assume that the users do not overload the system and therefore $S_i$ is smaller than or equal to
the number of frequency subcarriers, here two. Note that $\bP_i$ may have low row rank when certain subcarriers are not used. 
For example, if user $i$ 
transmits one symbol on subcarrier 1 but nothing on subcarrier 2, then $\bP_i= [a, 0; 0, 0]$ for some complex scalar $a$. 
 If $\bP_i$ is diagonal, then each 
symbol is only sent on one frequency.
Denote the $m$-th transmit symbol of user $i$ as $\x_i(m)$ which is randomly generated, mutually independent and with covariance matrix $\I_{2}$. 
The precoding matrix $\bP_i$ satisfies
the transmit power constraint of user $i$: $\tr\left( \bP_i \bP_i^{\her}\right)\leq P_i^{max}$.  
Denote the channel gain from transmitter (TX) $i$ to receiver (RX) $j$ on frequency $m$ as $h_{ji}(m)$. 
For simplicity of the example, we let $S_i$ equal two. The received 
signal of user $i$ is a vector whose $m$-th element is the received signal on the $m$-th frequency subcarrier,
\begin{equation}
 \y_i = \left[\begin{array}{c}
              \y_i(1)\\ \y_i(2)
             \end{array}
 \right]= \sum_{j=1}^2 \left[ \begin{array}{cc}
                               h_{ij}(1) & 0\\
			      0 & h_{ij}(2)
                              \end{array}
\right] \bP_j \left[ \begin{array}{c}
		   x_j(1)\\
		   x_j(2)
               \end{array}
\right] + \left[ \begin{array}{c}
        n_i(1)\\
	n_i(2)
                 \end{array}
\right].
\end{equation} The circular Gaussian noise with unit variance received on the $m$-th subcarrier at RX $i$ is denoted as $n_i(m)$.
If a relay with two antennas is introduced into the system, it receives the broadcasting signal from TXs and forwards 
them to RXs. We denote the received signal at the relay as a stacked vector of the received signal at each frequency $m$, with 
$\y_r(m)\in \bbC^{2 \times 1}$ representing the received signal on frequency $m$ and the $a$-th element in $\y_r(m)$ representing 
the signal at the $a$-th antenna:
\begin{equation}
 \y_r= \left[\begin{array}{c}
              \y_r(1)\\ \y_r(2)
             \end{array}
 \right]= \sum_{j=1}^2 \left[ \begin{array}{cc}
                  \f_j(1) & \0_{2 \times 1}\\
		  \0_{2 \times 1} & \f_j(2)
                 \end{array}
\right] \bP_j \left[ \begin{array}{c}
                x_j(1)\\
		x_j(2)
               \end{array}
\right]+ \left[ \begin{array}{c}
                 \n_r(1)\\
		  \n_r(2)
                \end{array}
\right]
\end{equation} where $\n_r(m)\in \bbC^{2 \times 1}$ is a circular Gaussian noise vector received at frequency $m$ with 
identity covariance matrix and $\f_j(m)$ is the complex vector channel from user $j$ to the relay on frequency $m$.
The relay processes the received signal $\y_r$ by a multiplication of matrix $\R \in \bbC^{4}$ and forwards the signal 
to the RXs. Denote the channel from relay to RX $i$ on frequency $m$ by $\g_i(m)\in \bbC^{2 \times 1}$. At RX $i$, the received 
signal is 
\begin{equation}\label{eqt:received_sig_ex}
\begin{aligned}
 \y_i& = \sum_{j=1}^2 \left( \left[ \begin{array}{cc}
                               h_{ij}(1) & 0\\
			      0 & h_{ij}(2)
                              \end{array}
\right]+ \left[ \begin{array}{cc}
                  \g^{\her}_i(1) & \0_{1 \times 2}\\
		  \0_{1 \times 2} & \g^{\her}_i(2)
                 \end{array}
\right] \R   \left[ \begin{array}{cc}
                  \f_j(1) & \0_{2 \times 1}\\
		  \0_{2 \times 1} & \f_j(2)
                 \end{array}
\right] \right) \bP_j \left[ \begin{array}{c}
                x_j(1)\\
		x_j(2)
               \end{array}
\right]\\
& + \left[ \begin{array}{cc}
                  \g^{\her}_i(1) & \0_{1 \times 2}\\
		  \0_{1 \times 2} & \g^{\her}_i(2)
                 \end{array}
\right] \R \left[ \begin{array}{c}
                 \n_r(1)\\
		  \n_r(2)
                \end{array}
\right]+ \left[ \begin{array}{c}
                     n_i(1)\\
		    n_i(2)
                    \end{array}
\right].
\end{aligned}
\end{equation}
Denote channel matrices
\begin{equation*}
 \bH_{ij}=\left[ \begin{array}{cc}
                               h_{ij}(1) & 0\\
			      0 & h_{ij}(2)
                              \end{array}
\right], \hspace{0.5cm} \G^{\her}_i=\left[ \begin{array}{cc}
                  \g^{\her}_i(1) & \0_{1 \times 2}\\
		  \0_{1 \times 2} & \g^{\her}_i(2)
                 \end{array}
\right], \hspace{0.5cm} \F_i= \left[ \begin{array}{cc}
                  \f_j(1) & \0_{2 \times 1}\\
		  \0_{2 \times 1} & \f_j(2)
                 \end{array}
\right] 
\end{equation*}
and the equivalent channel from TX $j$ to RX $i$ as
$
 \bar{\bH}_{ij}=\bH_{ij} + \G_i^{\her} \R \F_j.
$
 An achievable rate of user 1 is
\begin{equation}
 r_1(\R)= \cC\left( \I_2 + \bar{\bH}_{11} \bP_1 \bP_1^{\her} \bar{\bH}^{\her}_{11} 
\left( \bar{\bH}_{12} \bP_2 \bP_2^{\her} \bar{\bH}^{\her}_{12} + \G^{\her}_1 \R \R^{\her} \G_1 + \I_2\right)^{-1} \right).
\end{equation}
 Consider that RX $2$ is an eavesdropper. We compute the worst-case scenario in which RX $2$ decodes all other symbols perfectly 
before decoding the messages from TX $1$ and RX 2 sees a MIMO channel and decodes messages $x_1(1)$ and $x_2(2)$ utilizing 
both frequencies (with a MMSE receive filter for example).
\begin{equation}\label{eqt:in_leakage_ex}
\begin{aligned}
 \y_{2 \leftarrow 1}&=\left( \left[ \begin{array}{cc}
                               h_{21}(1) & 0\\
			      0 & h_{21}(2)
                              \end{array}
\right]+ \left[ \begin{array}{cc}
                  \g^{\her}_2(1) & \0_{1 \times 2}\\
		  \0_{1 \times 2} & \g^{\her}_2(2)
                 \end{array}
\right] \R   \left[ \begin{array}{cc}
                  \f_1(1) & \0_{2 \times 1}\\
		  \0_{2 \times 1} & \f_1(2)
                 \end{array}
\right] \right) \bP_1 \left[ \begin{array}{c}
                x_1(1)\\
		x_1(2)
               \end{array}
\right]\\
& + \left[ \begin{array}{cc}
                  \g^{\her}_2(1) & \0_{1 \times 2}\\
		  \0_{1 \times 2} & \g^{\her}_2(2)
                 \end{array}
\right] \R \left[ \begin{array}{c}
                 \n_r(1)\\
		  \n_r(2)
                \end{array}
\right]+ \left[ \begin{array}{c}
                     n_2(1)\\
		    n_2(2)
                    \end{array}
\right]
\end{aligned}
\end{equation} An achievable rate is then
$
r_{2 \leftarrow 1}(\R)= \cC \left( \I_2 + \bar{\bH}_{21} \bP_1 \bP_1^{\her} \bar{\bH}^{\her}_{21} \left( \G_2^{\her} \R \R^{\her} \G_2 + \I_2\right)^{-1}\right).
$
An achievable secrecy rate of user 1 is then the achievable rate of user 1 $r_1(\R)$ minus the leakage rate to user 2 $r_{2 \leftarrow 1}(\R)$ \cite{Khisti2010a}:
\begin{equation}\label{eqt:secrecy_rate_ex1}
\begin{aligned}
 r_1^{s}(\R)&= \left(r_1(\R) -r_{2 \leftarrow 1}(\R) \right)^{+}\\
&= \Bigg( \cC\left( \I_2 + \bar{\bH}_{11} \bP_1 \bP_1^{\her} \bar{\bH}^{\her}_{11} \left( \bar{\bH}_{12} \bP_2 \bP_2^{\her} \bar{\bH}^{\her}_{12} + \G^{\her}_1 \R \R^{\her} \G_1 + \I_2\right)^{-1} \right)\\
&-\cC \left( \I_2 + \bar{\bH}_{21} \bP_1 \bP_1^{\her} \bar{\bH}^{\her}_{21} \left( \G_2^{\her} \R \R^{\her} \G_2 + \I_2\right)^{-1}\right) \Bigg)^+.
\end{aligned}
\end{equation}

The relay processing matrix is defined as
\begin{equation}
 \R = \left[ \begin{array}{cc}
              \R_{11} & \R_{12}\\
	      \R_{21} & \R_{22}
             \end{array}
\right]
\end{equation} where each submatrix block $\R_{mn}$ forwards signals from frequency $n$ to frequency $m$. 
In a peaceful MIMO IRC, $\R$ bares a block diagonal structure, $\R_{12}=\R_{21}=\0_2$.
The intuition is that relays should not generate cross talk over frequency channels. However,
it is not trivial to examine the effect of $\R_{12}$ and $\R_{21}$ on secrecy rates as illustrated below and the conventional block diagonal 
structure should not be a-priori assumed.

\begin{table*}[!t]
\begin{center}
\caption{Randomly generated channel realizations for a two user two frequency interference relay channel with two antennas at relay and single antenna at TXs and RXs.
\label{table:channel}}
\begin{tabular}{|c|}
\hline
$\bH_{11} =\left[\begin{array}{cc}
    0.5129 + 0.4605i &        0\\
         0   & 0.3504 + 0.0950i
\end{array}\right], \hspace{0.1cm}
\bH_{21} =\left[\begin{array}{cc}
    0.4337 + 0.0709i &        0\\
         0  &  0.1160 + 0.0078i
\end{array}\right]$ \\
$\bH_{12} =\left[\begin{array}{cc}
    0.3693 + 0.0336i &        0\\
         0  &  0.1922 + 0.4714i
\end{array}\right], \hspace{0.1cm}
\bH_{22} =\left[\begin{array}{cc}
    0.1449 + 0.0718i  &       0\\
         0   & 0.6617 + 0.0432i
\end{array}\right]$\\
$ \G_1 =\left[\begin{array}{cc}
   0.4460 + 0.5281i   &     0   \\       
   0.5083 + 0.5729i   &     0    \\      
        0      &       0.3608 + 0.1733i\\
        0      &       0.3365 + 0.0861i
\end{array}\right], \hspace{0.1cm}
\G_2 =\left[\begin{array}{cc}
   0.3933 + 0.0111i    &    0         \\ 
   0.8044 + 0.2331i    &    0          \\
        0      &       0.9339 + 0.7859i\\
        0      &       0.2268 + 0.4107i
\end{array}\right]$ \\
$ \F_1 =\left[\begin{array}{cc}
   0.1194 + 0.8624i   &     0    \\      
   0.6344 + 0.1582i   &     0      \\    
        0        &     0.6012 + 0.6261i \\
        0        &     0.1176 + 0.8351i
\end{array}\right], \hspace{0.1cm}
\F_2 =\left[\begin{array}{cc}
   0.9404 + 0.2720i   &     0   \\       
   0.4156 + 0.9280i   &     0     \\     
        0      &       0.9213 + 0.8129i\\
        0      &       0.5420 + 0.1664i
\end{array}\right]$\\
\hline
$\R^{\IN} =\left[\begin{array}{cccc}
  -0.0364 - 0.0035i & -0.1793 - 0.0233i &  0.0234 - 0.0575i  & 0.0574 + 0.0596i\\
  -0.1046 + 0.0925i & -0.2837 - 0.0390i & -0.0832 - 0.0249i &  0.0029 + 0.1567i\\
   0.2729 + 0.0708i & -0.1376 + 0.1714i & -0.3130 - 0.2977i  & 0.2012 - 0.1606i\\
   0.0529 + 0.0099i & -0.1388 + 0.0348i & -0.4690 - 0.3154i & -0.0414 - 0.1751i\\
\end{array}\right]$\\
$\R^{\IN,d} =\left[\begin{array}{cccc}
   -0.0364 - 0.0035i & -0.1793 - 0.0233i &  0 &  0\\
   -0.1046 + 0.0925i & -0.2837 - 0.0390i  & 0  & 0\\
   0 &  0  & -0.3130 - 0.2977i  & 0.2012 - 0.1606i\\
   0 &  0  & -0.4690 - 0.3154i & -0.0414 - 0.1751i\\
\end{array}\right]$\\
$\R^{\IN,z}= \left[\begin{array}{cccc}
        -0.2709 + 0.2267i & -0.0820 + 0.1738i & -0.0770 + 0.0704i &  -0.1357 + 0.1183i\\
  -0.1509 + 0.0212i & -0.3225 - 0.4885i & -0.2088 - 0.0485i &  0.6810 + 0.1046i\\
   0.2459 + 0.1223i & -0.1315 + 0.0682i & -0.2702 - 0.2781i &  0.2683 - 0.2842i\\
  -0.0155 + 0.1640i & -0.2285 - 0.0472i & -0.5114 - 0.2436i & -0.0346 - 0.1960i \\         
\end{array}\right]$\\
\hline
\end{tabular}
\vspace{0.2cm}

\end{center}
\end{table*}

As a numerical example, we compute the secrecy rates with the following randomly generated channels given in Table \ref{table:channel}. We set the
precoding matrices of TX 1 and TX 2 to be
\begin{equation*}
 \bP_1=\left[ \begin{array}{cc}
               1& 0\\
		1&0
              \end{array}
\right], \hspace{0.3cm} \bP_2=\left[ \begin{array}{cc} 1&4\\-4 & 1 \end{array}\right]
\end{equation*} which means that TX 1 transmits only one data stream on both subcarriers and TX 2 transmits two
data streams spread over both frequency subcarriers with orthogonal sequences. With relay matrix $\R^{\IN}$ (see Table \ref{table:channel})
a sum secrecy rate of 3.4104 is achievable whereas with block diagonal matrix $\R^{\IN,d}$ the sum secrecy rate is 3.1881.
A block diagonal relay matrix does not always improve secrecy rate and therefore in the following we assume a general non-block-diagonal structure $\R$. In fact, the relay matrix $\R^{\IN}$ is chosen such that the secrecy leakage is zero: $(\bH_{12} + \G_1^{\her} \R \F_2) \bP_2=\0$ and 
$(\bH_{21} + \G_2^{\her} \R \F_1) \bP_1=\0$. 
Thus, the secrecy rate from \eqref{eqt:secrecy_rate_ex1} can be simplified to the following
\begin{equation}\label{eqt:ach_rate_in_ex2}
  r_i^{s} = \cC \left( \I_2 + \bar{\bH}_{11} \bP_1 \bP_1^{\her} \bar{\bH}^{\her}_{11} \left(  \G^{\her}_1 \R \R^{\her} \G_1 + \I_2\right)^{-1} \right).
\end{equation}
This motivates our following proposition on information leakage neutralization techniques. Interestingly, with information leakage 
neutralization, we can simplify the optimization problem significantly. The idea is to set the 
information leakage from each user at each frequency to zero, in particular, by setting the equivalent 
channel of $\x_1$ from TX 1 to RX 2 and vice versa in \eqref{eqt:in_leakage_ex} to zero,
\begin{equation}\label{eqt:in_ex_block}
\left\{
\begin{aligned}
& \left(\bH_{12}+ \G_1^{\her} \R \F_2 \right) \bP_2 = \0\\
& \left(\bH_{21}+ \G_2^{\her} \R \F_1 \right) \bP_1 = \0.\\
\end{aligned}
\right. 
\end{equation}
With the properties of the Kronecker product, \eqref{eqt:in_ex_block} can be written as 
\begin{equation}
\left[
 \begin{array}{c}
  \left( \left(\F_2 \bP_2 \right)^{\tran} \otimes \G_1^{\her} \right)\\
\left( \left(\F_1 \bP_1 \right)^{\tran} \otimes \G_2^{\her} \right)
\end{array} \right] \bvec(\R) = \B \bvec(\R)= \left[\begin{array}{c}
-\bvec(\bH_{12} \bP_2)\\
-\bvec(\bH_{21} \bP_1)\\
 \end{array} \right] = \bb
\end{equation}
The stacked matrix $\B$ in the above equation is a fat matrix\footnote{Care must be taken when users send less than $M$ data streams
 (when $\bP_i$ has zero columns. More discussion is provided later in Proposition 2).}. We obtain the relay matrix that can perform information leakage neutralization:
\begin{equation}\label{eqt:r_i^sn_num1}
 \bvec(\R)= \B^{\her} \left( \B \B^{\her} \right)^{-1} \bb.
\end{equation}
Substitute the channel realizations in Table \ref{table:channel} into the above equation and reverse the vectorization operation, 
we obtain the relay matrix $\R^{\IN}$ (please refer to the table for numerical values). 
\begin{Remark}
 If the precoding matrices $\{\bP_i\}$ are invertible, then the relay matrix $\R$ obtained using \eqref{eqt:r_i^sn_num1} is block diagonal.
A block diagonal relay matrix means that the relay sets cross talk over frequency subcarriers to zero and due to the interference
leakage neutralization, the interference from users on the same frequency is also zero. This results in $KM$ parallel channels without 
interference. We propose in Section \ref{sec:effin} a suboptimal but very efficient algorithm which optimizes the achievable rates in 
this case\footnote{The achievable rates here are secrecy rates as the information leakage is zero.}. 
\end{Remark}
In fact, the matrix in \eqref{eqt:r_i^sn_num1}
 is not unique, any matrix which is a sum of $\bvec(\R)$ in \eqref{eqt:r_i^sn_num1} and
a vector in the null space of $\B$ can also neutralize information leakage,
\begin{equation}\label{eqt:r_i^sn_num2}
 \bvec(\R)= \B^{\her} \left( \B \B^{\her} \right)^{-1} \bb + \z,
\end{equation} where $\z \in \mathcal{N}(\B)$. With the channel realizations given in Table \ref{table:channel}, 
we can generate another matrix $\R^{\IN,z}$ which achieves a higher secrecy rate 4.1553, a 17.8\% increase of secrecy 
rate by optimization over $\z$. This motivates us to investigate an efficient method to find $\z$ and consequently $\R$ 
which neutralizes information leakage and optimizes the secrecy rate at the same time. 
\begin{Remark}
 With the optimization over $\z$, the relay matrix is no longer block diagonal which couples the frequency channels. Although
the problem is more complicated, we have shown in the above example that one can get a better secrecy rate performance. 
In Section \ref{sec:optin}, we propose an iterative sum secrecy rates optimization over the relay matrix $\R$ and the precoding matrices $\{\bP_i\}$.
\end{Remark}
In the following section, we illustrate how the relay matrix can 
be chosen carefully to amplify the desired signal strength and at the same time neutralize information leakage 
in the multi-user scenario.

\section{General multi-user multi-antenna multi-carrier scenario}
In this section, we let the number of TXs and RXs be $K\geq2$. The TXs and RXs have single antenna and 
the relay  has $N$ antennas. Let the number of frequency subcarriers be $M$.
Denote the complex channel from TX $i$ to RX $j$, as a diagonal matrix $\bH_{ji} \in \bbC^{M}$ and the complex channel  from 
TX $i$ to relay as $\F_{i} \in \bbC^{NM \times M}$ and from relay to RX $j$ as $\G_{j} \in \bbC^{MN \times M}$. The signal received at the relay is,
\begin{equation}
 \mathbf{y}_r=  \sum_{i=1}^K \F_{i} \bP_i\x_i+ \mathbf{n}_r
\end{equation}
where $\F_i=\diag\left(\f_i(1), \ldots, \f_i(M) \right)$ and
$\x_i \in \bbC^{M \times 1}$  are the circular Gaussian transmit symbols from TX $i$, with zero mean and identity covariance matrix.
The matrix $\bP_i\in \bbC^{M}$ satisfies the power constraint:
\begin{equation}
 \tr \left( \bP_i \bP_i^{\her} \right) \leq P_i^{max}.
\end{equation}
With AF strategy, the relay multiplies the received signal $\y_r$ on the left by processing matrix $\R$ and transmits $\R \y_r$.
The transmit power of the relay is constrained by $P_r^{max}$,
\begin{equation}\label{eqt:pow_constraint}
 \tr\left( \R \left( \sum_{i=1}^K \F_{i} \bP_i \bP_i^{\her} \F_{i}^{\her} + \I_{\mn} \right) \R^H\right) \leq P_r^{max}.
\end{equation}
The received signal at RX $j$ is 
\begin{equation}\label{eqt:in_out}
 \y_j= \sum_{i=1}^K \left( \bH_{ji} + \G_{j}^{\her} \R \F_{i}\right) \bP_i \x_i + \G_{j}^{\her} \R \mathbf{n}_r + \n_j
\end{equation} where $\n_j$ is the circular Gaussian noise at RX $j$ with zero mean and identity covariance matrix and
$\G_{j}=\diag(\g_j(1), \ldots, \g_j(M))$.
For the ease of notation, we define the equivalent channel from $i$ to $j$ as
\begin{equation}\label{eqt:equiv}
 \bar{\bH}_{ji}=\bH_{ji} + \G_{j}^H \R \F_{i}
\end{equation} and its $(f,m)$-element is $[\bar{\bH}_{ji}]_{fm}=h_{ji}+ \g_j^{\her}(f) \R_{fm} \f_i(m)$ which is the equivalent channel 
from user $i$ frequency $m$ to user $j$ frequency $f$.

Each RX is not only interested in decoding its own signal but also eavesdropping from other TXs. In the following,
 we define the worst case achievable secrecy rate  with colluding eavesdroppers.
 For messages $\x_i$, 
all RXs except RX $i$ collaborate to form an eavesdropper with multiple antennas and the message $\x_i$ goes 
through a multi-carrier MIMO channel to the  colluding eavesdroppers. A worst case secrecy rate is then to assume that  
all other messages $\x_j, j\neq i$ are decoded perfectly and subtracted before decoding $\x_i$.
The received signals at RX $i$ and the colluding eavesdroppers are
\begin{equation}\label{eqt:in_out_eave}\left\{
 \begin{aligned}
  \y_i &=\sum_{k=1}^K \bar{\bH}_{ik} \bP_k \x_k  + \G^{\her}_{i} \R \mathbf{n}_r + \n_i\\
  \y_{-i}&= \left[ \begin{array}{c}
				  \bar{\bH}_{1i}\\
				  \vdots\\
				   \bar{\bH}_{(i-1)i} \\
 \bar{\bH}_{(i+1)i}\\ \vdots\\ 
\bar{\bH}_{Ki} 
                               \end{array}
\right] \bP_i \x_i +
\left[ \begin{array}{c}
 \G_{1}^{\her} \\
  \vdots\\
\G_{i-1}^{\her} \\
\G_{i+1}^{\her} \\ \vdots\\ 
\G_{k}^{\her}
 \end{array}
\right] \R \n_r+
\left[ \begin{array}{c}
\n_1\\
\vdots\\
\n_{i-1}\\
\n_{i+1}\\
\vdots\\
\n_{K}        
       \end{array}
\right]\\
&= \bar{\bH}_{-i} \bP_i \x_i + \G_{-i}^{\her} \R \n_r + \n_{-i}.
 \end{aligned} \right.
\end{equation}
The secrecy rate of user $i$ is \cite{Khisti2010a},
\begin{equation}\label{eqt:secrecy_rate}
\begin{aligned}
 r_i^s&=  \Bigg( \cC\left( \I_{M} + \bar{\bH}_{ii} \bP_i \bP_i^{\her} \bar{\bH}^{\her}_{ii} 
\left( \sum_{j \neq i} \bar{\bH}_{ij} \bP_j \bP_j^{\her} \bar{\bH}^{\her}_{ij} + \G^{\her}_{i}\R \R^{\her} \G_{i} + \I_{M}\right)^{-1} \right) \\
&- \cC \bigg( \I_{M(K-1)} + \bar{\bH}_{-i} \bP_i \bP_i^{\her} \bar{\bH}^{\her}_{-i} \left( \G^{\her}_{-i}\R \R^{\her} \G_{-i} + \I_{M(K-1)}\right)^{-1} \bigg) \Bigg)^+ .
\end{aligned}
\end{equation} 
Recall from \eqref{eqt:equiv} that the equivalent channel from Tx $j$ to Rx $i$ $\bar{\bH}_{ij}$ is a function of the relay 
processing matrix $\R$, $\bar{\bH}_{ij}=\bH_{ij} + \G_i^{\her} \R \F_j$. The optimization of the aforementioned 
secrecy rates is highly complicated due to their non-convex structure. In the following, we propose the information leakage neutralization 
 technique \cite{Ho2011d} which is able to neutralize 
all information leakage to all eavesdroppers in the air by choosing the relay strategy in a careful manner. 
As illustrated in the previous section, with information leakage neutralization, the secrecy rate expression \eqref{eqt:secrecy_rate}
can be simplified to
\begin{equation}\label{eqt:secrecy_rate_multiuser}
 r_i^s=\cC\left( \I_{M} + \bar{\bH}_{ii} \bP_i \bP_i^{\her} \bar{\bH}^{\her}_{ii} 
\left( \G^{\her}_{i}\R \R^{\her} \G_{i} + \I_{M}\right)^{-1} \right).
\end{equation}
In the following section, we illustrate how we can choose $\R$ to achieve a secrecy rate as such.

\subsection{Information Leakage Neutralization}\label{sec:in}
We choose $\R$ such that the equivalent channel of message $\x_i$ to the eavesdropper in 
\eqref{eqt:in_out_eave} is neutralized to zero. The challenge of information leakage neutralization in multi-subcarrier environment as compared to 
the single-subcarrier case \cite{Ho2011d} is that the information leakage neutralization constraints must be modified 
to incorporate frequency sharing:
\begin{equation}\label{eqt:in_const}
 \left(\bH_{ji} + \G_j^{\her} \R \F_i \right) \bP_i=0, \hspace{1cm} i,j=1,\ldots, K, i \neq j.
\end{equation}
Note that we 
consider the most general scenario where users may only use part of the spectrum  and send less than $M$ data streams and 
thus $\bP_i$ may have zero rows and zero columns.
In the following, we show the dependency of the number of antennas at the relay for information leakage neutralization
on these system parameters.
\PropBox{
\label{prop:in_dim}
The number of antennas at the relay, $N$, required to neutralize all information leakage from each of the $K$ users at each frequency subcarrier, in
a total of $M$ subcarriers, satisfies
\begin{equation}
 N \geq \sqrt{\frac{K-1}{M}\sum_{i=1}^K S_i }
\end{equation}
where $S_i$ is the number of data streams sent by TX $i$ .
}
For the proof, please refer to Appendix \ref{app:relay_in_num_ant}.
Proposition 1 offers the minimum number of antennas required to ensure secrecy which
depends on the number of users $K$, the number subcarriers $M$ and the amount of data streams transmitted $S_i$.
\begin{itemize}
\item If every user employs full frequency multiplexing $S_i=M$, we have then
\begin{equation}
 N \geq \sqrt{\frac{K-1}{M}\sum_{i=1}^K M}= \sqrt{K(K-1)}.
\end{equation} As $N$ is an integer, we have $N \geq K$ which is the same criteria as in the flat-fading case \cite{Ho2011d}. 
\item If every user sends $S_i=aM$ data streams and $0 \leq a \leq 1$, we have then
\begin{equation}
 N \geq \sqrt{\frac{K-1}{M} \sum_{i=1}^K a M} = \sqrt{a K(K-1)}.
\end{equation}
For example, in a scenario of $K=3$ users, $M=16$ frequency subcarriers and each user transmits
$S_i=8$ data streams $\left(a=\frac{1}{2}\right)$, 
the relay must have at least $\left\lceil \sqrt{ \frac{1}{2} \cdot 3 \cdot 2} \right\rceil= 
\left\lceil \sqrt{3} \right\rceil=2$ 
antennas to completely remove any information leakage from any TX to any RX. This is  
less than $\lceil \sqrt{3(2)} \rceil=3$ if all users send $S_i=M=16$ data streams.
\item Note that the number of antennas required for information leakage neutralization is \emph{independent}
to the number of frequency subcarriers used by each user (the number of non-zero rows of $\bP_i$)\footnote{The reason is that even if a user does not transmit
on a certain frequency, the relay must make sure that it does not forward the user's information on other subcarriers to 
this subcarrier at which the eavesdroppers can decode the information.}. However, the power required to neutralize information leakage depends on
how crowded the subcarriers is. If
a lot of frequency subcarriers are occupied, the relay may not have enough power to neutralize all information leakage as we will see
in the following.
\end{itemize}
When the number of antennas at the relay is sufficient for information leakage neutralization, we can 
use the following method to compute the relay forwarding matrix $\R$ for such purpose.

\PropBox{
 Any relay matrix $\R$ satisfying the information leakage neutralization constraint \eqref{eqt:in_const1} has the following form:
\begin{equation*}
 \bvec(\R)= \A^{\dagger} \bb + \z
\end{equation*}
where 
\begin{equation*}
\begin{aligned}
 \A&=\left[\left( \left( \hat{\bP}^{\tran}_1 \F_1^{\tran} \right) \otimes \G^{\her}_{-1} \right)^{\her}, 
\ldots, \left( \left( \hat{\bP}^{\tran}_K \F_K^{\tran} \right) \otimes \G^{\her}_{-K} \right)^{\her} \right]^{\her}\\
\bb&= \left[ -\bvec\left(\bH_{-1} \hat{\bP}_1\right)^{\her}, \ldots, 
-\bvec\left( \bH_{-K} \hat{\bP}_K\right)^{\her}\right]^{\her}\\
\z&\in \nnull\left( \A \right).
\end{aligned}
\end{equation*} and $\hat{\bP}_i$ is a submatrix of $\bP_i$, containing its non-zero columns.
}
\vspace{0.2cm}
For the proof, please refer to Appendix \ref{app:relay_in_form}. From Proposition 2, 
it follows that there is a minimum power requirement for information leakage neutralization. 

\begin{Corollary}\label{cor:min_power}
The minimum power required for information leakage neutralization is
\begin{equation*}
 P_r^{max} \geq \left(\A^{\dagger} \bb \right)^{\her} \left( \left( \sum_{i=1}^K \F_{i} \bP_i \bP_i^{\her} \F_{i}^{\her} + \I_{\mn} \right)
\otimes \I_{\mn} \right) \left(\A^{\dagger} \bb\right).
\end{equation*}
\vspace{0.05cm}
\end{Corollary}
For  the proof, please refer to Appendix \ref{app:min_power}.
Depending on the available transmit power at the relay, one may only have enough power to neutralize information leakage but not enough
power to further improve the transmission rates. If there is  limited power resource and therefore
one must ensure secure transmission with as little power as possible, then one can set $\z$ in Proposition 2 to zero. If there is a
high priority of secrecy rates and with abundant transmit power, one can optimize $\z$ for the purpose of sum secrecy rate maximization. 
In the following, we investigate algorithms to address these applications.

\section{Information Leakage Neutralization Algorithms }\label{sec:propose}
In the previous section, we have shown that secrecy rates \eqref{eqt:secrecy_rate_multiuser} are achievable by information leakage neutralization. Also,
in order to implement information leakage neutralization, the number of antennas at the relay, the number of frequency subcarriers and the number of users
in the system must satisfy the relation in Proposition 1. In Proposition 2, we computed the minimum relay power required in order to perform information 
leakage neutralization. With more power available at the relay, we can improve the achievable secrecy rates by optimizing the relay matrix and 
the precoding matrices. The optimization of sum
secrecy rates can be written formally in the following:
\boxeqn{
\begin{aligned}
\max_{\R,\{ \bP_i\}} \hspace{0.5cm} & \sum_{i=1}^{K} \cC\left( \I_{M} + \bar{\bH}_{ii} \bP_i \bP_i^{\her} \bar{\bH}^{\her}_{ii} 
\left( \G^{\her}_{i}\R \R^{\her} \G_{i} + \I_{M}\right)^{-1} \right)\\
\st \hspace{0.5cm} & \tr \left( \bP_i \bP_i^{\her} \right) \leq P_i^{max}\\
& \tr\left(\R \left(\sum_{i=1}^K \F_i \bP_i \bP_i^{\her} \F_i^{\her} \right) \R^{\her} \right) \leq P_r^{max}.
\end{aligned}
}

In the following, we propose two algorithms. The first algorithm EFFIN, in Section \ref{sec:effin}, considers the scenario where $\z=\0$ in Proposition 2
 and all users transmit
the maximum number of data streams allowed $S_i=M$. We observe that in this situation, information leakage neutralization 
decomposes the system into $KM$ parallel channels and consequently both the relay processing matrix $\R$ and the precoding matrix $\bP_i$
can be computed very efficiently. The second algorithm OPTIN, in Section \ref{sec:optin}, investigates a systematic method for the computation of
$\R$ and $\bP_i$ when there is enough transmit power budget at the relay to allow further optimization of secrecy rates.

\subsection{Efficient Information Leakage Neutralization (EFFIN)}\label{sec:effin}
When every user transmits $S_i=M$ data streams and $\bP_i$ is invertible, we propose the following algorithm that decomposes the 
$K$ user interference relay channels with $M$ frequency subcarriers and $N$ antennas at the relay to $KM$ parallel secure channels 
\emph{with no interference and no information leakage}.
The information leakage neutralization criteria
$
 \left( \bH_{ij} + \G_{i}^{\her} \R \F_j \right) \bP_i = \0,
$ when $\bP_i$ is invertible, is equivalent to
\begin{equation*}
 \bH_{ij} + \G_{i}^{\her} \R \F_j  = \0.
\end{equation*} Due to the block diagonal structure of $\bH_{ij}$, $\G_i$ and $\F_j$, one feasible solution of the above equation is a block diagonal $\R$.
With the block diagonal structure, the resulting secrecy rates may be suboptimal, but the information leakage neutralization constraint
can be broken down to the optimization over the diagonal blocks $\R_{mm}$ in $\R$:
\begin{equation}
 h_{ji}(m)+ \g_j^{\her}(m) \R_{mm} \f_i(m)=0, \hspace{0.3cm} i,j=1,\ldots,K, i\neq j.
\end{equation} Following the same approach as before, we stack the constraints for all $j \neq i$ and define
\begin{equation*}
\begin{aligned}
 \h_{-i}(m)&=\left[h^{\her}_{1i}(m), \ldots, h^{\her}_{(i-1)i}(m), h^{\her}_{(i+1)i}(m),\ldots, h^{\her}_{Ki}(m)\right]^{\her}\\
\G_{-i}(m)&=\left[\g_1(m),\ldots,\g_{i-1}(m),\g_{i+1}(m),\ldots, \g_{K}(m) \right].
\end{aligned}
\end{equation*} We obtain $\h_{-i}(m) + \G^{\her}_{-i}(m) \R_{mm} \f_i(m)=\0_{(K-1) \times 1}$ which is equivalent to
\begin{equation*}
 \left(\f_i^{\tran}(m) \otimes \G^{\her}_{-i}(m)\right) \bvec\left(\R_{mm} \right)= - \h_{-i}(m).
\end{equation*} Stacking constraints for all $i$, we have
\begin{equation*}
 \A(m)=\left[\begin{array}{c}
             \left(\f_1^{\tran}(m) \otimes \G^{\her}_{-1}(m)\right) \\
\vdots\\
\left(\f_K^{\tran}(m) \otimes \G^{\her}_{-K}(m)\right)
             \end{array} \right], \hspace{0.5cm}
\bb(m)= \left[\begin{array}{c}
           - \h_{-1}(m)  \\
\vdots\\
- \h_{-K}(m)  
              \end{array}
 \right].
\end{equation*} With a limited power budget at relay, we propose to implement information leakage neutralization with
the least relay transmit power and utilize the result from Proposition 2, the relay matrix has the $m$-th diagonal block equal to
\begin{equation} \label{eqt:block_diag_R}
 \R_{mm}= \bvec^{-1}\left(\left(\A(m) \right)^{\dagger} \bb(m)\right)
\end{equation} where $\bvec(.)^{-1}$ is to reverse the vectorization of a vector columnwise to a $M \times M$ matrix.
After the computation of the relay matrix in \eqref{eqt:block_diag_R}, $\R=\diag\left( \R_{11},\ldots, \R_{MM}\right)$, 
the optimal precoding matrices $\{\bP_i\}$ are computed by solving $\cQ_1$.
\boxeqn{
\begin{aligned}
  \cQ_1: \hspace{0.1cm} \max_{\{ \Q_i \}, \Q_i \succeq 0} \hspace{0.3cm} & \sum_{i=1}^K \cC \left( \I_M +   \Q_i \W_i \right)\\
\st  \hspace{0.3cm} & \tr \left( \Q_i \right) \leq P_i^{max}, \hspace{0.2cm} i=1,\ldots, K,\\
&  \sum_{i=1}^K \tr \left( \Q_i \X_i \right) \leq \bar{P}_r^{max}.
\end{aligned}
}
where we replace $\bP_i \bP_i^{\her}$ by positive semi-definite variable $\Q_i$ and denote the following matrices
\begin{equation}
\begin{aligned}
 \W_i&=\left( \bH_{ii}+ \G_i^{\her}\R \F_i \right)^{\her} \left( \G_i^{\her} \R \R^{\her} \G_i+ \I_M\right)^{-1} \left( \bH_{ii}+ \G_i^{\her}\R \F_i \right),\\
\X_i&=\F_{i}^{\her}\R^{\her} \R \F_{i}, \\
\bar{P}_r^{max}&= P_r^{max}- \tr \left( \R \R^{\her}\right).
\end{aligned}
\end{equation}
The objective in $\cQ_1$
is concave in  $\Q_i$ as $\W_i$ is positive semi-definite and the constraints are linear in $\Q_i$. 
Thus, $\cQ_1$ is a semi-definite program and can be solved readily using convex optimization solvers, e.g. CVX%
\footnote{Given block diagonal $\R$ in \eqref{eqt:block_diag_R}, the equivalent channel $\W_i$ and matrix $\X_i$ are also 
block diagonal. It is possible to solve $\cQ_1$ using water-filling with $K+1$ Lagrange multipliers. For large problem size,
it may be more computational efficient using a tailor made water-filling method. For medium size problems and illustrative purposes, 
we propose here to solve by semi-definite programming. }. The optimal $\bP_i$ is obtained by performing eigenvalue 
decomposition on $\Q_i= \U_i \D_i \U^{\her}_i$ and $\bP_i=\U_i \D_i^{1/2}$.
The psuedocode of the EFFIN is given in Algorithm \ref{algo:effin}.

\begin{algorithm}
\caption{The pseudo-code for Efficient Information Leakage Neutralization (EFFIN) \label{algo:effin}}
\begin{algorithmic}[1]
  \For{$m=1 \to M$}  \Comment{Compute block diagonal relay processing matrix } 
  \State Compute $\R_{mm}=\bvec^{-1}\left(\left(\A(m) \right)^{\dagger} \bb(m)\right)$ with 
\begin{equation*}
 \A(m)=\left[\begin{array}{c}
             \left(\f_1^{\tran}(m) \otimes \G^{\her}_{-1}(m)\right) \\
\vdots\\
\left(\f_K^{\tran}(m) \otimes \G^{\her}_{-K}(m)\right)
             \end{array} \right], \hspace{0.5cm}
\bb(m)= \left[\begin{array}{c}
           -\h_{-1}(m)  \\
\vdots\\
-\h_{-K}(m)  
              \end{array}
 \right].
\end{equation*} 
  \EndFor
\State The relay processing matrix is $\R=\diag\left(\R_{11},\ldots, \R_{MM} \right)$.  
  \State Solve $\cQ_1$ using convex optimization solvers and obtain optimal $\{\Q_i\}$.  
  \For{$i=1 \to K$}\Comment{Compute precoding matrices}
  \State Perform eigen-value decomposition, $\Q_i=\U_i \D_i \U_i^{\her}$. Set $\bP_i=\U_i \D_i^{1/2}$.
  \EndFor
\end{algorithmic}
\end{algorithm}

\subsection{Optimized Information Leakage Neutralization (OPTIN)}\label{sec:optin}
In the previous subsection, we have discussed a simple, efficient and power saving solution of the relay matrix and precoding matrices for secure transmission. 
One drawback of the efficient method is that its performance may be suboptimal. In this subsection, we discuss how to choose the relay and precoding matrices 
such that the sum secrecy rates are optimized while ensuring zero information leakage.

To this end, we rewrite the information leakage neutralization constraint \eqref{eqt:in_const} to promote the optimization of secrecy rates, 
\begin{equation}\label{eqt:in_block}
 \left(\bH + \G^{\her} \R \F \right) \bP = \T
\end{equation} where $\bH=\left[\bH_{11}, \ldots, \bH_{1K}; \ldots; \bH_{K1},\ldots, \bH_{KK} \right]$, $\G^{\her}=\left[ \G_1^{\her}; \ldots; \G_K^{\her}\right]$,
$\F=\left[ \F_1, \ldots, \F_K\right]$ and $\bP=\diag(\bP_1,\ldots, \bP_K)$. The block diagonal matrix $\T=\diag(\T_1,\ldots, \T_K)$ is 
the new optimization variable. $\T_i$ is the equivalent desired channel from TX $i$ to RX $i$ as $\T_i= (\bH_{ii} + \G^{\her}_i \R \F_i) \bP_i$.
By applying pseudo-inverses
\footnote{Note that $\G^{\her}$ 
has dimension $MK \times MN$ and $\F \bP$ has dimension $MN \times KM$. If $MN\geq MK$, 
then $\G^{\her \dagger}= \G \left( \G^{\her} \G\right)^{-1}$ and 
$ \left( \F \bP \right)^{\dagger}=\left( \left( \F \bP \right)^{\her}\left( \F \bP \right)\right)^{-1}\left( \F \bP \right)^{\her}$. 
If $MN< KM$, 
then $\G^{\her \dagger}=\left(\G \G^{\her}\right)^{-1}\G$ and 
$\left( \F \bP \right)^{\dagger}=\left(\F \bP\right)^{\her} \left( \F \bP \left(\F \bP\right)^{\her}\right)^{-1}$.}
 of $\G^{\her}$ and $\F \bP$ ($\G^{\her \dagger}$ and $\left(\F \bP \right)^\dagger$ respectively),
 one can rewrite \eqref{eqt:in_block} to the following
\begin{equation}\label{eqt:r_i^sntermsof_t}
 \R= \G^{\her \dagger} \left( \T- \bH \bP\right) \left( \F \bP \right)^{\dagger}.
\end{equation}
The maximum achievable sum secrecy rate is the solution of the following problem
\begin{subequations}
\begin{align}
  \max_{ \R, \T, \{\bP_i \}} \hspace{0.3cm} & 
 \sum_{i=1}^K \cC \left( \I_M +  \T_i \bP_i \bP_i^{\her} \T_i^{\her} \left( \G_i^{\her} \R \R^{\her} \G_i+ \I_M\right)^{-1} \right)\\
\st \hspace{0.3cm} &  \tr \left( \bP_i \bP_i^{\her}\right) \leq P_i^{max}, \hspace{0.2cm} i=1, \ldots, K, \label{opt:trans_pow_constraint}\\
& \left(\bH + \G^{\her} \R \F \right) \bP = \T, \label{opt:in_constraint}\\
&  \tr \left( \R \left( \F \bP \bP^{\her} \F^{\her} + \I_{\mn}\right) \R^{\her} \right) \leq P_r^{max}\label{opt:relay_pow_constraint}\\
& \T=\diag\left( \T_1,\ldots, \T_K\right). 
\end{align}
\end{subequations}
Note that in the objective function, the information leakage is neutralized for each user. 
Constraints \eqref{opt:trans_pow_constraint} and \eqref{opt:relay_pow_constraint} are 
the transmit power constraints at the TXs and at the relay respectively. The information leakage neutralization constraint is written as \eqref{opt:in_constraint}.
The optimization is not jointly convex in $\R$, $\T$ and $\{\bP_i\}$. 
To simplify the optimization problem, 
we propose the following iterative optimization algorithm. Given $\R$ and $\T$, we solve $\bP_i$ optimally using $\cQ_1$ in EFFIN.

The second part of the iterative algorithm is to compute the optimal relay strategy $\R$ and the auxiliary variable $\T$ (by solving $\cQ_2$)
if the precoding matrices $\bP_i$ as the solutions of $\cQ_1$ are given.
\boxeqn{
\begin{aligned}
  \cQ_2: \hspace{0.3cm} \max_{ \R, \T} \hspace{0.3cm} & \sum_{i=1}^K \cC \left( \I_M +  \T_i \T_i^{\her} \left( \G_i^{\her} 
\R \R^{\her} \G_i+ \I_M\right)^{-1} \right)\\
\st \hspace{0.3cm} & \R= \G^{\her \dagger} \left( \T- \bH \bP\right) \left( \F \bP \right)^{\dagger},\\
& \tr \left( \R \left( \F \bP \bP^{\her} \F^{\her} + \I_{\mn}\right) \R^{\her} \right) \leq P_r^{max},\\
& \T=\diag(\T_1,\ldots, \T_K).
 \end{aligned}
}
Problem $\cQ_2$ is non-convex. The major challenge is due to the sum of log-determinants in the objective function and 
the equality constraints. 
In the following, we utilize the first equality constraint and replace $\R$ as a function of $\T$. 
The optimization problem $\cQ_2$ can be written as, 
\boxeqn{
\begin{aligned}
  \cQ_2': \hspace{0.3cm} \max_{ \T} \hspace{0.3cm} & \sum_{i=1}^K \left( \cC \left( \X_i + \bar{\T}_i \Z_i \bar{\T}_i^H \right)
-\cC \left( \X_i + \bar{\T}_i \Y_i \bar{\T}_i^{\her} \right) \right)\\
\st \hspace{0.3cm}  & \tr \left( \G^{\her \dagger} \left( \T- \bH \bP\right) \left( \tilde{\F}+ \I_{MK}\right) 
\left( \T- \bH \bP\right)^{\her}\G^{ \dagger}\right)\leq P_r^{max},\\
& \bar{\T}_i =\left[ \T_i, \I_M\right],\\
& \T=\diag(\T_1,\ldots, \T_K).
 \end{aligned}
}
Please see the proof and the definition of $\X_i, \Y_i, \Z_i$ in \eqref{eqt:xyz} in Appendix \ref{app:cq2}. 
Although the optimization problem is simplified, it is still non-convex in $\T$. In the following, we propose to solve $\cQ_2'$ with gradient descent method. 
To this end, we write the Lagrangian of $\cQ_2'$ as $L(\T,\lambda)$,
\begin{equation}\label{eqt:lagrangian}
 \begin{aligned}
  L(\T,\lambda)&=\sum_{i=1}^K \left( \cC \left( \X_i + \bar{\T}_i \Z_i \bar{\T}_i^H \right)
-\cC \left( \X_i + \bar{\T}_i \Y_i \bar{\T}_i^{\her} \right) \right)\\
& -\lambda \left( \tr \left( \G^{\her \dagger} \left( \T- \bH \bP\right) \left( \tilde{\F}+ \I_{MK}\right)
\left( \T- \bH \bP\right)^{\her}\G^{ \dagger}\right)- P_r^{max} \right)\\
&= \sum_{i=1}^K f_i(\T_i) - \lambda g(\T).
 \end{aligned}
\end{equation} 
The gradient of the Lagrangian with respect to $\T^*$ is
\begin{equation}\label{eqt:gradient}
  \begin{aligned}
 \cD_{\T^*} L(\T,\lambda)
&= \frac{1}{\ln(2)} \left[ \begin{array}{cccc}
                            \cD_{\T_1^*} f_1(\T_1) & \0_M & \ldots & \0_M\\
\0_M & \cD_{\T_2^*} f_2(\T_2) & \ldots & \0_M\\
& &  \ddots & \vdots\\
\0_M &\ldots &  & \cD_{\T_K^*} f_K(\T_K)
                           \end{array}
 \right]\\
& - \lambda \G^{ \dagger}\G^{\her \dagger} \left( \T- \bH \bP\right) 
\left( \tilde{\F}+ \I_{KM}\right).
\end{aligned}
\end{equation} Please see the proof in Appendix \ref{app:lagrangian}.
We summarize in Algorithm \ref{algo:optin} the proposed iterative algorithm on sum secrecy rate optimization.
\begin{algorithm}
\caption{The pseudo-code for Optimized Information Leakage Neutralization (OPTIN)\label{algo:optin}}
\begin{algorithmic}[1]
  \While{}\Comment{Compute relay processing matrix }
  \State Initialize $\{\bP_i\}$  and $\R$ as the solutions of EFFIN.
  \State Solve $\cQ_2'$ using gradient descent method with gradient \eqref{eqt:gradient} and obtain optimal solution $\T$.
Obtain relay processing matrix $\R$ from $\T$ using \eqref{eqt:r_i^sntermsof_t}. 
  \State With $\R$ and $\T$ above, solve $\cQ_1$ using convex optimization solvers and obtain optimal $\{\Q_i\}$.  
  \For{$i=1 \to K$}\Comment{Compute precoding matrices}
  \State Perform eigen-value decomposition, $\Q_i=\U_i \D_i \U_i^{\her}$. Set $\bP_i=\U_i \D_i^{1/2}$.
  \EndFor
  \If{ sum secrecy rate improvement is less than a predefined threshold}
      \State Convergence reached. Break.
  \EndIf
 \EndWhile
\end{algorithmic}
\end{algorithm}

\section{Simulation Results}\label{sec:sim}
To illustrate the effectiveness of the proposed algorithms, we provide in this section numerical simulations for different system settings. As an 
example, we simulate the secrecy rates of a relay assisted network with $K=2$ users, $M=8$ frequency subcarriers and $N=2$ antennas at the relay, 
unless otherwise stated.
To examine the performance of the algorithms with respect to system signal-to-noise ratio, we vary the transmit power constraint at relay from $0$ to $30 \dB$
while keeping the transmit power constraint at TXs as $10 \dB$ (see Figure \ref{fig:relay_pow}.) Similarly, we examine the algorithms by varying
the transmit power constraint at TXs from $0$ to $30 \dB$ while keeping the transmit power at relay constraint at $23, 27,30 \dB$. Note that by varying the 
power constraints, we do not force the power of the optimized precoding matrices and the relay processing matrix to be equal to the power constraints.
In the following, we compare algorithms:
\begin{itemize}
 \item Baseline 1 (Repeater): the relay is a layer 1 relay and is only able to forward signals without additional signal processing. This corresponds to
setting $\R=\I_{MN} \sqrt{\frac{P_r^{max}}{MN}}$.
 \item Baseline 2 (IC): the relay shuts down, i.e. $\R=\0_{MN}$, and we obtain an interference channel where users eavesdrop each other.
 \item Proposed algorithm EFFIN: an efficient relay and precoding matrices optimization algorithm outlined in Algorithm \ref{algo:effin}.
 \item Proposed algorithm OPTIN: an optimized algorithm whose performance exceeds EFFIN with a price of higher complexity. OPTIN is outlined in 
Algorithm \ref{algo:optin}. 
\end{itemize}
For each baseline algorithm, we examine the effect of spectrum sharing on achievable secrecy rates by employing either one of the following spectrum
sharing methods:
\begin{itemize}
 \item Full spectrum sharing (FS): users are allowed to use the entire spectrum. Each TX measures the channel qualities of the direct channel and 
the channel from itself to other RXs. Based on the measured channel qualities, each TX excludes frequency subcarriers with zero secrecy rates and transmits
on the channels with non-zero secrecy rates. For subcarriers at which more than one user would like to transmit, we assume that the TXs coordinate
so that the TX with a high secrecy rate would transmit on that subcarrier. Despite such coordination, each user eavesdrops other users on each subcarrier.
 \item Orthogonal spectrum sharing (OS): users are assigned exclusive portion of spectrum. Each TX excludes subcarriers with zero secrecy rates and  
transmits on the channels with non-zero secrecy rates. Each user eavesdrops other users on each subcarrier.
\end{itemize}

\subsection{Secrecy rates with increasing relay power}
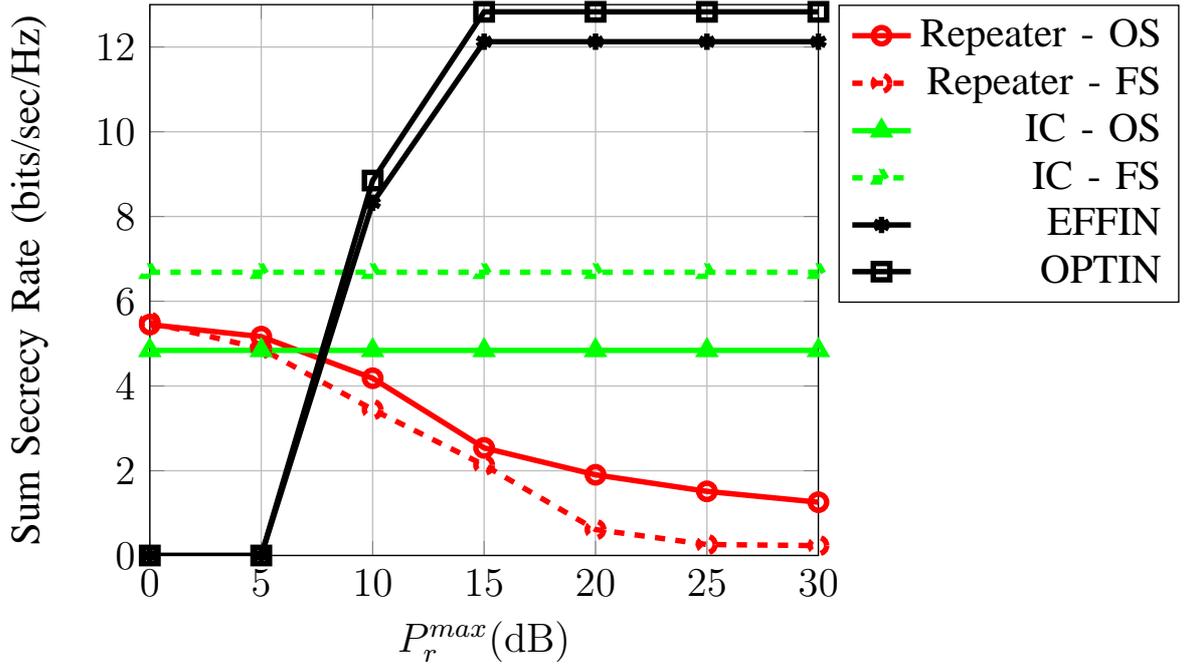
\begin{figure}
 \begin{center}
\resizebox{\linewidth}{!}{
\begin{tikzpicture}
\pgfplotsset{every axis legend/.append style={
at={(0.97,0.4)},
anchor=east}}

\begin{axis}[xmin=0,xmax=30,ymin=0,ymax=13,
ylabel={Sum Secrecy Rate (bits/sec/Hz)},
xlabel={$P_r^{max} (\dB)$}, 
legend style={
cells={anchor=east},
legend pos=outer north east,
},
grid=major]

\tikzstyle{every pin}=[font=\footnotesize]
\tikzstyle{every mark}=[scale=1.2]

\addplot[ultra thick, red, mark=o, solid] coordinates {
(0,5.4492)    
(5, 5.1647)
(10,4.1832)
(15, 2.5393)
(20, 1.9047)
(25, 1.5136)
(30, 1.2576)
};
 \addlegendentry{Repeater - OS};

\addplot[ultra thick,red,mark=o,dashed] coordinates {
(0, 5.5027)    
(5, 4.8820)
(10, 3.4429)
(15, 2.1336)
(20, 0.6122)
(25, 0.2590)
(30, 0.2335)            
}; 
\addlegendentry{Repeater - FS};

\addplot[ultra thick,green,mark=triangle,solid] coordinates {
(0,   4.8409)    
(5,   4.8409)
(10,   4.8409)
(15,   4.8409)
(20,   4.8409)
(25,  4.8409)
(30,   4.8409)  
};
 \addlegendentry{IC - OS};

\addplot[ultra thick,green,mark=triangle,dashed] coordinates {
(0,   6.6850)    
(5,   6.6850)
(10,   6.6850)
(15,  6.6850)
(20,  6.6850)
(25,  6.6850)
(30, 6.6850)  
};
 \addlegendentry{IC - FS};

\addplot[ultra thick,black,mark=asterisk,solid] coordinates {
(0,  0)    
(5,  0)
(10,  8.3232)
(15,  12.1250)
(20,  12.1250)
(25,  12.1250)
(30,  12.1250)  
};
 \addlegendentry{EFFIN};

\addplot[ultra thick,black,mark=square,solid] coordinates {
(0,  0)    
(5,  0)
(10,  8.8481)
(15,  12.83)
(20,  12.83)
(25,  12.83)
(30,  12.83)  
};
 \addlegendentry{OPTIN};

\end{axis}
\end{tikzpicture}
} 
\caption{The achievable secrecy rates of a two-user IRC with 8 frequency subcarriers is shown with varying relay power constraint. 
The TX power constraints are $10 \dB$ and there
are two antennas at the relay. The proposed scheme EFFIN and OPTIN outperform baseline algorithms Repeater and IC by 550\% and 200\% respectively.}
\label{fig:relay_pow}
 \end{center}
\end{figure}

In Figure \ref{fig:relay_pow}, we show  achievable sum secrecy rates over varying the transmit power constraint at the relay from $0$ to $30 \dB$
while keeping the transmit power constraint at  the TXs at $10 \dB$. As the IC does not
utilize the relay, the achievable sum secrecy rates (plotted with triangles) are constant as the
relay power constraint increases. As expected from intuition, the performance of IC with
FS is better than OS because OS has an additional constraint of subcarrier assignment. 
The achievable sum secrecy rates achieved by a repeater decreases with relay transmit power.
This is due to the increased amplification noise in AF relaying. Interestingly, the non-intelligent
relaying scheme, e.g. a repeater, may decrease the secrecy rate significantly, even worse than switching
off the relay. However, utilizing an intelligent relay and choosing the relaying scheme, one
can improve the achievable secrecy rate significantly, about 550\% over a simple repeater and about 200\% 
over IC. Although EFFIN is very simple and efficient, it achieves 94.5\% of the sum secrecy rate achieved by the  
more complicated algorithm OPTIN.

\subsection{Secrecy rates with increasing TX power}
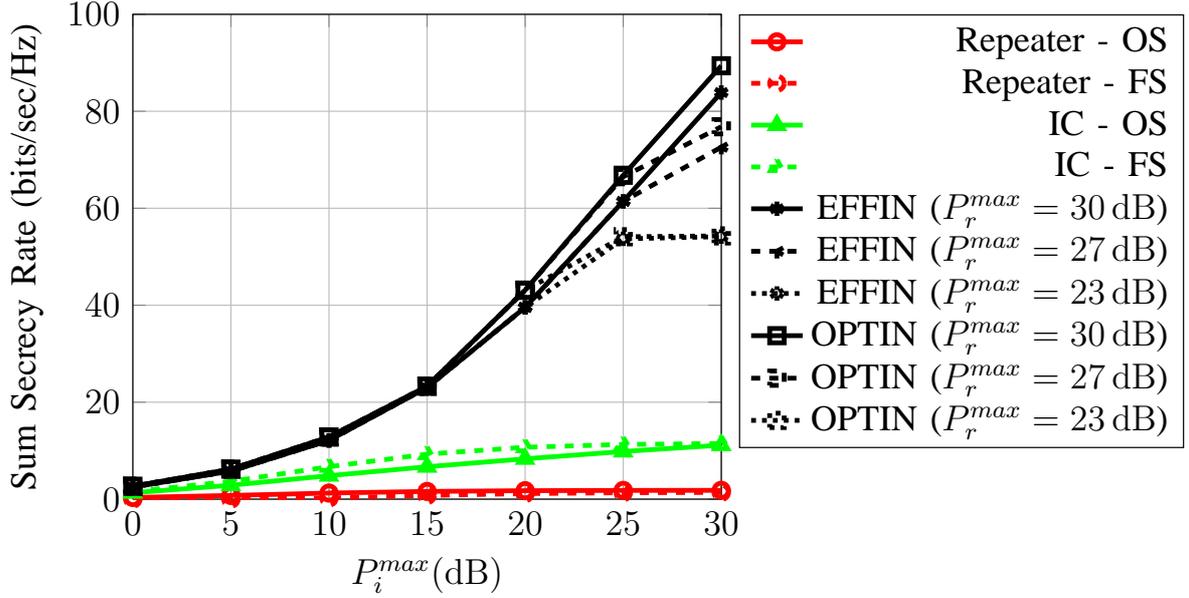
\begin{figure}
 \begin{center}
\resizebox{\linewidth}{!}{
\begin{tikzpicture}
\pgfplotsset{every axis legend/.append style={
at={(0.97,0.4)},
anchor=east}}

\begin{axis}[xmin=0,xmax=30,ymin=0,ymax=100,
ylabel={Sum Secrecy Rate (bits/sec/Hz)},
xlabel={$P_i^{max} (\dB)$}, 
legend style={
cells={anchor=east},
legend pos=outer north east,
},
grid=major]

\tikzstyle{every pin}=[font=\footnotesize]
\tikzstyle{every mark}=[scale=1.2]

\addplot[ultra thick, red, mark=o, solid] coordinates {
(0,0.3490)    
(5, 0.7671)
(10,1.2576)
(15, 1.5948)
(20, 1.7482)
(25, 1.8040)
(30, 1.8225)                     
};
 \addlegendentry{Repeater - OS};

\addplot[ultra thick,red,mark=o,dashed] coordinates {
(0, 0)    
(5, 0)
(10, 0.2335)
(15, 0.7633)
(20, 1.1494)
(25,1.3275)
(30, 1.3926)            
}; 
\addlegendentry{Repeater - FS};

\addplot[ultra thick,green,mark=triangle,solid] coordinates {
(0,    1.3208)    
(5,   2.8706)
(10,  4.8409)
(15,  6.6831)
(20,  8.3177)
(25,  9.8314 )
(30,  11.1372)  
};
 \addlegendentry{IC - OS};

\addplot[ultra thick,green,mark=triangle,dashed] coordinates {
(0,  1.5629  )    
(5,  3.6811 )
(10,  6.6850)
(15,  9.2851)
(20,  10.7083)
(25,  11.2800)
(30,  11.4774)  
};
 \addlegendentry{IC - FS};

\addplot[ultra thick,black,mark=asterisk,solid] coordinates {
(0,  2.4896)    
(5,   5.7870)
(10,  12.1250 )
(15,  23.0302)
(20,  39.5633 )
(25,  61.4396)
(30,  83.8611)  
};
 \addlegendentry{EFFIN ($P_r^{max}=30\dB$)};

\addplot[ultra thick,black,mark=asterisk,dashed] coordinates {
(0,   2.4896)    
(5,   5.7870)
(10,  12.1250 )
(15,  23.0302)
(20,  39.5633 )
(25,  61.4340)
(30,  72.5904)  
};
 \addlegendentry{EFFIN ($P_r^{max}=27\dB$)};

\addplot[ultra thick,black,mark=asterisk,dotted] coordinates {
(0,  2.4896)    
(5,   5.7870)
(10,  12.1250 )
(15,  23.0302)
(20,  39.5633 )
(25,  53.6883)
(30,  54.1439)  
};
 \addlegendentry{EFFIN ($P_r^{max}=23\dB$)};

\addplot[ultra thick,black,mark=square,solid] coordinates {
(0,  2.667)    
(5,  6.196 )
(10,  12.83)
(15,  23.29)
(20,  43.1)
(25,  66.8)
(30,  89.39)  
};
 \addlegendentry{OPTIN ($P_r^{max}=30\dB$)};

\addplot[ultra thick,black,mark=square,dashed] coordinates {
(0,    2.6659 )    
(5,  6.1978 )
(10,  12.8237)
(15,  23.29)
(20,  43.1)
(25,  66.5052)
(30,  76.8086 )  
};
 \addlegendentry{OPTIN ($P_r^{max}=27\dB$)};

\addplot[ultra thick,black,mark=square,dotted] coordinates {
(0,  2.667)    
(5,  6.196 )
(10,  12.83)
(15,  23.29)
(20,  43.1)
(25,  54.12)
(30,  54.3)  
};
 \addlegendentry{OPTIN ($P_r^{max}=23\dB$)};
\end{axis}
\end{tikzpicture}
} 
\caption{The achievable secrecy rates of a two-user IRC with 8 frequency subcarriers is shown with varying transmitter power constraints. 
The relay power constraint is $30 \dB$ and there
are two antennas at the relay. The secrecy rates achieved by EFFIN and OPTIN grows unbounded with the transmit power at TX whereas
the secrecy rates achieved by baseline algorithms saturate in high SNR regime.}
\label{fig:tx_pow}
 \end{center}
\end{figure}

In Figure \ref{fig:tx_pow}, we simulate the achievable sum secrecy rate by the transmit power constraint at TXs from 
$0$ to $30 \dB$ while keeping the transmit power at relay constraint at $23, 27, 30 \dB$. As the transmit power at TX
increases, the sum secrecy rates saturate in both baseline algorithms, Repeater and IC. With the proposed information
leakage neutralization, we see that the sum secrecy rates grow unbounded with the TX power as each user enjoys 
a leakage free frequency channel. Note that the sum secrecy rates achieved by relay with power constraint at $23, 27, 30 \dB$ are
plotted in dotted, dashed and solid lines respectively.  When there is only $23 \dB$ available, there is only enough power 
for information leakage neutralization, but
not enough to further optimize the system performance. Hence, the achievable sum secrecy rates of EFFIN and OPTIN overlap. 
With more power available, it is possible to optimize the sum secrecy rates while neutralizing information leakage and the performance
of OPTIN is better than EFFIN.  

\subsection{Secrecy rates with larger systems}
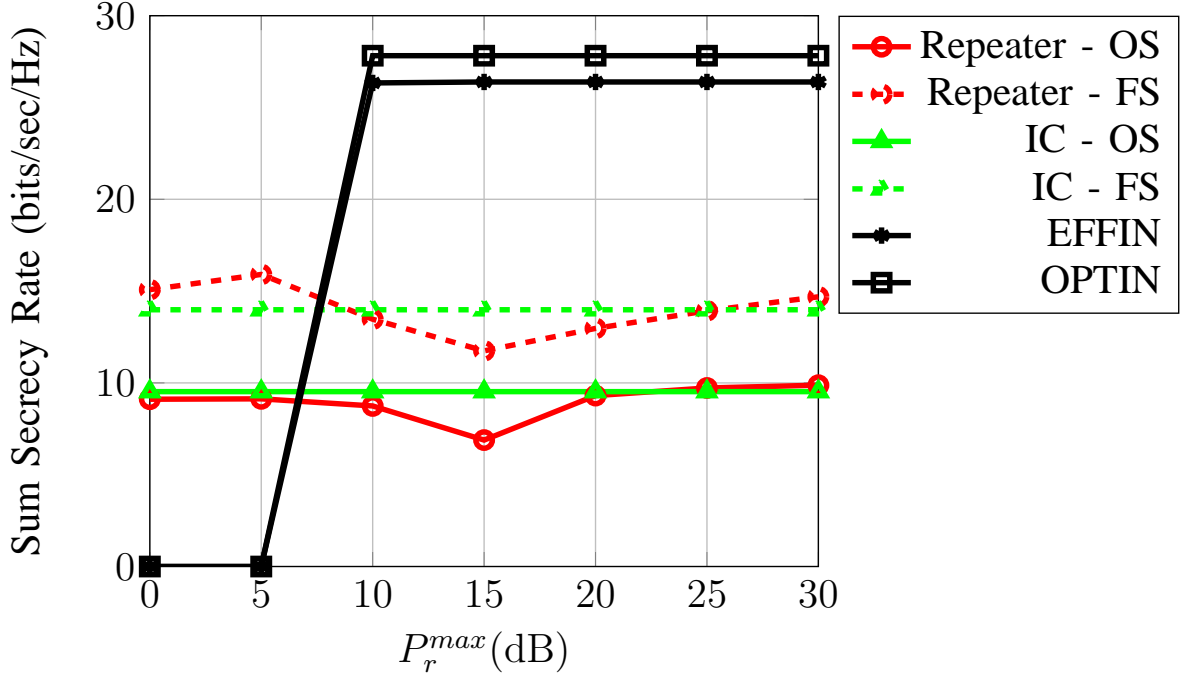
\begin{figure}
 \begin{center}
\resizebox{\linewidth}{!}{
\begin{tikzpicture}
\pgfplotsset{every axis legend/.append style={
at={(0.97,0.4)},
anchor=east}}

\begin{axis}[xmin=0,xmax=30,ymin=0,ymax=30,
ylabel={Sum Secrecy Rate (bits/sec/Hz)},
xlabel={$P_r^{max} (\dB)$}, 
legend style={
cells={anchor=east},
legend pos=outer north east,
},
grid=major]

\tikzstyle{every pin}=[font=\footnotesize]
\tikzstyle{every mark}=[scale=1.2]

\addplot[ultra thick, red, mark=o, solid] coordinates {
(0,9.1108)    
(5, 9.1321)
(10,8.7451)
(15, 6.8866)
(20,  9.3093 )
(25, 9.7238)
(30, 9.8758)                     
};
 \addlegendentry{Repeater - OS};
                        
\addplot[ultra thick,red,mark=o,dashed] coordinates {
(0, 15.0778)    
(5, 15.9084 )
(10, 13.4685)
(15, 11.7447)
(20, 12.9683 )
(25,13.9265)
(30, 14.6895)            
}; 
\addlegendentry{Repeater - FS};

\addplot[ultra thick,green,mark=triangle,solid] coordinates {
(0,    9.5247)    
(5,   9.5247)
(10,  9.5247)
(15,  9.5247)
(20,  9.5247)
(25, 9.5247 )
(30,  9.5247)  
};
 \addlegendentry{IC - OS};

\addplot[ultra thick,green,mark=triangle,dashed] coordinates {
(0,  13.9837 )    
(5,  13.9837 )
(10,  13.9837)
(15,  13.9837)
(20,  13.9837)
(25,  13.9837)
(30,  13.9837)  
};
 \addlegendentry{IC - FS};

\addplot[ultra thick,black,mark=asterisk,solid] coordinates {
(0,  0)    
(5,   0)
(10,  26.3367 )
(15,  26.3893 )
(20,  26.3893)
(25,  26.3893)
(30,  26.3893)  
};
 \addlegendentry{EFFIN};

\addplot[ultra thick,black,mark=square,solid] coordinates {
(0,  0)    
(5,  0)
(10,  27.8201)
(15,  27.8201)
(20,  27.8201)
(25,  27.8201)
(30,  27.8201)  
};
 \addlegendentry{OPTIN};

\end{axis}
\end{tikzpicture}
} 
\caption{The achievable sum secrecy rates of a two-user IRC with 16 frequency subcarriers and 4 antennas at the relay is shown with varying  
relay power constraint. 
The TX power constraints are $10 \dB$ and there
are two antennas at the relay. The proposed scheme EFFIN and OPTIN outperform baseline algorithms Repeater and IC by 200\%. EFFIN achieves 94.86\%
of the sum secrecy rate performance by OPTIN.}
\label{fig:sim_txpow_m16}
 \end{center}
\end{figure}

In Figure \ref{fig:sim_txpow_m16}, we examine the performance of the proposed algorithms in a slightly larger systems with
$N=4$ antennas at the relay and $M=16$ frequency subcarriers. The relay processing matrix is therefore a $64 \times 64$ matrix.
The proposed scheme EFFIN and OPTIN outperform baseline algorithms Repeater and IC by 200\% whereas the efficient EFFIN algorithm achieves 94.86\%
of the sum  secrecy rate performance by OPTIN.

\appendices
\section{Proof of Proposition 1}\label{app:relay_in_num_ant}
If TX $i$ transmits $S_i\leq M$ data streams, then $M-S_i$ columns
of $\bP_i$ are zeros. For example, in a system with 4 subcarriers where TX $i$ transmits 2 data streams spread over 3 subcarriers, 
$\bP_i$ has the following form,
\begin{equation}
 \bP_i= \left[ \begin{array}{cccc}
                \ast & \ast & 0 & 0\\
\ast & \ast&  0 & 0\\
0 & 0 & 0 & 0\\
\ast & \ast& 0 & 0
               \end{array}
\right].
\end{equation}
Denote the non-zero columns of $\bP_i$ by $\hat{\bP}_i \in \bbC^{M \times S_i}$. The information 
leakage constraint \eqref{eqt:in_const} is equivalent to
\begin{equation}\label{eqt:in_const1}
 \left(\bH_{ji} + \G_j^{\her} \R \F_i \right) \hat{\bP}_i=0, \hspace{1cm} i,j=1,\ldots, K, i \neq j.
\end{equation}
For each $i$, we stack the constrains for all $j \neq i$ by using $\G^{\her}_{-i}$ from \eqref{eqt:in_out_eave} and defining 
\begin{equation*}
  \bH_{-i}=[\bH^{\her}_{1i}, \ldots, \bH^{\her}_{(i-1)i}, \bH^{\her}_{(i+1)i}, 
\ldots, \bH^{\her}_{Ki} ]^{\her}.
\end{equation*}
We write \eqref{eqt:in_const1} as
\begin{equation}
 \left(\bH_{-i}+ \G^{\her}_{-i} \R \F_i\right) \hat{\bP}_i =0, \hspace{1cm} i=1,\ldots, K
\end{equation}
which can be manipulated to the following by performing vectorization on the matrices,
\begin{equation}\label{eqt:in_const2}
  \left( \left( \hat{\bP}^{\tran}_i \F_i^{\tran} \right) \otimes \G^{\her}_{-i} \right) \bvec(\R)
= - \bvec \left(\bH_{-i} \hat{\bP}_i\right), \hspace{0.2cm} i=1, \ldots, K.
\end{equation}
The matrix $\bH_{-i}$ has dimension $(K-1) M \times M$ and the matrix $\hat{\bP}_i$ has dimension $M \times S_i$. 
Hence, the product $\bH_{-i} \bP_i$ has dimension $(K-1)M \times S_i$. The number of 
constraints in \eqref{eqt:in_const2} 
is the number of elements in $\bH_{-i} \bP_i$, which is $(K-1)M S_i$. Summing up all constraints for $i=1,\ldots,K$, 
we have the total number of constraints $(K-1)M \sum_{i=1}^K S_i $. The number of variables is the number of elements in 
$\R$ which equals to $M^2 N^2$. To neutralize information leakage at all users, we must satisfy \eqref{eqt:in_const2} 
for all $i$. To this end, the relay must have the number of antennas $N$ satisfying $ M^2 N^2 \geq (K-1)M\sum_{i=1}^K S_i$, or
\begin{equation}
 N \geq  \sqrt{\frac{K-1}{M}\sum_{i=1}^K S_i }.
\end{equation}

\section{Proof of Proposition 2}\label{app:relay_in_form}
 Stacking the matrices in \eqref{eqt:in_const1} for all $i$, we obtain  $\A \bvec(\R)= \bb$. 
The matrix $\A$ is a block matrix with vertically stacked blocks 
$\left( \hat{\bP}^{\tran}_i \F_i^{\tran} \right) \otimes \G^{\her}_{-i}$ , for $i=1, \ldots, K$, and therefore
has dimension $\sum_{i=1}^{K} S_i (K-1)M \times M^2 N^2$.  
The matrix $\G_{-i}$ concatenates matrices $\G_j$ for $j \neq i$, e.g., $\G_{-1}=[\G_2, \ldots, \G_K]$. 
As $\G_{-i}$ are not mutually independent, $\A$ is of low rank. Denote the number of rows of $\A$ by 
$\alpha=\sum_{i=1}^{K} S_i (K-1)M$ and the rank of $\A$ by $\beta=\rank(\A)$.
The pseudo-inverse of $\A$ can be computed by performing singular-value-decomposition on $\A$,
\begin{equation}
\begin{aligned}
& \left[\A\right]_{\alpha \times M^2 N^2}\\
&= \left[ \U_1 | \U_2 \right] \left[\begin{array}{cc}
                                       \vGamma & \0_{\beta \times (M^2 N^2-\beta)}\\
\0_{(\alpha-\beta) \times \beta} & \0_{(\alpha-\beta) \times (M^2 N^2-\beta)}
                                      \end{array}
 \right] \left[ \begin{array}{c}
       \V_1^{\her}\\
	\V_2^{\her}
                \end{array}
\right],
\end{aligned}
\end{equation}
where $\U_1 \in \bbC^{\alpha \times \beta}, \U_2 \in \bbC^{\alpha \times (\alpha-\beta)}$ are the left singular vectors in 
the signal space and null space of $\A$ respectively; $\V_1^{\her} \in \bbC^{\beta \times M^2 N^2}$, $\V_2^{\her} \in \bbC^{(M^2 N^2-\beta) \times M^2 N^2}$ 
are the right singular vectors in the signal space and null space of $\A$ respectively; $\vGamma \in \bbC^{\beta \times \beta}$ holds the 
non-zero singular values in the diagonal and zeros everywhere else. Thus, the solution of $\bvec(\R)$ satisfying $\A \bvec(\R)=\bb$ is
\begin{equation}\label{eqt:R_vec}
 \bvec(\R)= \V_1 \vGamma^{-1} \U_1^{\her}\bb + \V_2 \y
\end{equation} where $\y$ is any vector in the space of $\bbC^{M^2 N^2 \times 1}$. The result follows by setting $\z=\V_2 \y$ as a vector
in the null space of $\A$.

\section{Proof of Corollary \ref{cor:min_power}}\label{app:min_power}
Using the properties of Kronecker products, the relay transmit power from \eqref{eqt:pow_constraint} is equivalent to
$\left(\A^{\dagger} \bb +\z\right)^{\her} \left( \left( \sum_{i=1}^K \F_{i} \bP_i \bP_i^{\her} \F_{i}^{\her} + \I_{\mn} \right)
\otimes \I_{\mn} \right) \left(\A^{\dagger} \bb+\z\right)$.
By Proposition 2 and \eqref{eqt:pow_constraint}, the minimum transmit power required to satisfy information leakage neutralization is 
\begin{equation}
\begin{aligned}
& \min_{\z} \left(\A^{\dagger} \bb +\z\right)^{\her} \left( \left( \sum_{i=1}^K \F_{i} \bP_i \bP_i^{\her} \F_{i}^{\her} + \I_{\mn} \right)
\otimes \I_{\mn} \right) \left(\A^{\dagger} \bb+\z\right)\\
& \xLeftrightarrow[]{\z=\0}  \tr\left( \left(\A^{\dagger} \bb \right)\left( \sum_{i=1}^K \F_{i} \bP_i \bP_i^{\her} \F_{i}^{\her} + \I_{\mn} \right)
\left(\A^{\dagger} \bb\right)^{\her} \right)\leq P_r^{max},
\end{aligned}
\end{equation} where the transition is due to the fact that $\z$ is in the null space of $\A$ and
 the fact that $\Q= \left( \sum_{i=1}^K \F_{i} \bP_i \bP_i^{\her} \F_{i}^{\her} + \I_{\mn} \right)
\otimes \I_{\mn} $ is positive semi-definite and $\z^{\her} \Q \z \geq 0$ for any $\z$.

\section{Formulation of $\cQ_2'$}\label{app:cq2}
Let $\E_i^T= \e_i^T \otimes \I_M$, $\bar{\T}_i=[\T_i, \I_M]$ and
\begin{equation}\label{eqt:xyz}
\begin{aligned}
 \tilde{\F}&=\left( \F \bP \right)^{\dagger}\left( \F \bP \right)^{\her \dagger}, 
 \X_i=\sum_{m=1}^K \sum_{l=1}^K \bH_{im} \bP_m \tilde{\F}_{ml} \bP_l^{\her} \bH_{il}^{\her},\\
\Y_i&=\left[ \begin{array}{cc}
                                    \tilde{\F}_{ii} & -\sum_{l=1}^K \tilde{\F}_{il} \bP_l^{\her} \bH_{il}^H\\
-\sum_{m=1}^K \bH_{im} \bP_m \tilde{\F}_{mi} & \I_M
                                   \end{array}
\right], \\
\Z_i&= \left[ \begin{array}{cc}
            \I_{M} & \0_{M}   \\
\0_{M}& \0_M 
             \end{array}\right] + \Y_i.
\end{aligned}
\end{equation}
With the equality constraint \eqref{eqt:in_block}, the amplification noise can be written as \eqref{eqt:amp_noise1}
\begin{figure*}[!h]
\begin{equation}\label{eqt:amp_noise1}
\begin{aligned}
 &\G_i^{\her} \R \R^{\her} \G_i
 = \E_i^{\tran}\G^{\her} \R \R^{\her} \G \E_i\\
&= \E_i^{\tran} \left( \T- \bH \bP\right) \left( \F \bP \right)^{\dagger}\left( \F \bP \right)^{\her \dagger}\left( \T- \bH \bP\right)^{\her} \E_i\\
&= \left[- \bH_{i1} \bP_1 , \ldots, \T_i - \bH_{ii} \bP_{i}, \ldots, -\bH_{iK} \bP_K \right] \left[\begin{array}{ccc}
                                                                                 \tilde{\F}_{11} &\ldots & \tilde{\F}_{1K}\\
\vdots & \ddots & \vdots\\
\tilde{\F}_{K1} & \ldots & \tilde{\F}_{KK}
                                                                                \end{array}
 \right] \left[ \begin{array}{c}
                 - \bP_1^{\her} \bH_{i1}^{\her}\\
\vdots\\
\T_i^{\her}- \bP_i^{\her} \bH_{ii}^{\her}\\
\vdots\\
-\bP_K^{\her} \bH_{iK}^{\her}
                \end{array}
\right]\\
&= \sum_{m=1}^K \sum_{l=1}^K \bH_{im} \bP_m \tilde{\F}_{ml} \bP_l^{\her} \bH_{il}^{\her}- \T_i \sum_{l=1}^K \tilde{\F}_{il} \bP_l^{\her} \bH_{il}^H
 -\sum_{m=1}^K \bH_{im} \bP_m \tilde{\F}_{mi}\T_i^{\her} + \T_i \tilde{\F}_{ii} \T_i^{\her}\\
&= \X_i - \I_M + \left[\T_i, \I_M \right] \left[ \begin{array}{cc}
                                    \tilde{\F}_{ii} & -\sum_{l=1}^K \tilde{\F}_{il} \bP_l^{\her} \bH_{il}^H\\
-\sum_{m=1}^K \bH_{im} \bP_m \tilde{\F}_{mi} & \I_M
                                   \end{array}
\right] \left[ \begin{array}{c}
                \T_i^H\\
		\I_M
               \end{array}
\right]\\
&= \X_i - \I_M + \bar{\T}_i \Y_i \bar{\T}_i^{\her},
\end{aligned}
\end{equation}
\hrule
\begin{equation}\label{eqt:pow_const1}
 \begin{aligned}
  &\tr\left(\R \left(\F \bP \bP^{\her} \F^{\her}+ \I_{\mn}\right) \R^{\her}\right) \\
&=\tr \left( \G^{\her \dagger} \left( \T- \bH \bP\right) \left( \F \bP \right)^{\dagger} \left(\F \bP \bP^{\her} \F^{\her}+ 
\I_{\mn}\right)\left( \F \bP \right)^{\dagger \her}\left( \T- \bH \bP\right)^{\her}\G^{ \dagger}\right)\\
&=\tr \left( \G^{\her \dagger} \left( \T- \bH \bP \right) \left(  \left( \F \bP \right)^{\dagger} \left( \F \bP \right)^{\dagger \her} + 
\I_{KM} \right)\left( \T- \bH \bP \right)^{\her}\G^{ \dagger}\right)\\
&=\tr \left( \G^{\her \dagger} \left( \T- \bH \bP\right) \left( \tilde{\F}+ \I_{KM} \right)\left( \T- \bH \bP\right)^{\her}\G^{ \dagger}\right)\leq P_r^{max}.
 \end{aligned}
\end{equation}
\hrule
\end{figure*}
%
%
%
where 
 $\tilde{\F}_{ml}\in \bbC^{M}$ is the $(m,l)$-th block matrix in $\tilde{\F}$. 
As a result, the objective can be written as
\begin{equation*}
 \begin{aligned}
 & \sum_{i=1}^K \cC \left( \I_M +  \T_i \T_i^{\her} \left( \G_i^{\her} \R \R^{\her} \G_i+ \I_M\right)^{-1} \right)\\
&= \sum_{i=1}^K \Bigg( \cC \left( \I_M +  \T_i \T_i^{\her} + \G_i^{\her} \R \R^{\her} \G_i\right)\\
& \hspace{0.5cm} - \cC \left( \I_M + \G_i^{\her} \R \R^{\her} \G_i \right) \Bigg)\\
&= \sum_{i=1}^K \Bigg( \cC \left( \X_i + \bar{\T}_i \Z_i \bar{\T}_i^H \right)
-\cC \left( \X_i + \bar{\T}_i \Y_i \bar{\T}_i^{\her} \right) \Bigg).
 \end{aligned}
\end{equation*} 
Similarly, the power constraint is written as \eqref{eqt:pow_const1}.

\section{Computation of the gradient of Lagrangian \eqref{eqt:lagrangian} }\label{app:lagrangian}
Recall the Lagrangian from \eqref{eqt:lagrangian},
\begin{equation*}
 \begin{aligned}
 & L(\T,\lambda)=\sum_{i=1}^K \left( \cC \left( \X_{i} + \bar{\T}_i \Z_{i} \bar{\T}_i^H \right)
-\cC \left( \X_i + \bar{\T}_i \Y_{i} \bar{\T}_i^{\her} \right) \right)\\
& -\lambda \left( \tr \left( \G^{\her \dagger} \left( \T- \bH \bP\right) \left( \tilde{\F}+ \I_{MK}\right)
\left( \T- \bH \bP\right)^{\her}\G^{ \dagger}\right)- P_r^{max} \right)\\
&= \sum_{i=1}^{K} f_i(\T_i)- \lambda g(\T),
 \end{aligned}
\end{equation*} where $f_i(\T_i)= \cC \left( \X_{i} + \bar{\T}_i \Z_{i} \bar{\T}_i^H \right)
-\cC \left( \X_i + \bar{\T}_i \Y_{i} \bar{\T}_i^{\her} \right) $ denotes the secrecy rates of TX \nolinebreak$i$ and $g(\T)$ denotes the power constraint. 
We compute the gradient of the Lagrangian \eqref{eqt:lagrangian} with respect to $\T$,
\begin{equation*}
 \cD_{\T^*}L(\T,\lambda)= \cD_{\T^*} \sum_{i=1}^K f_i(\T_i)-\lambda \cD_{\T^*} g(\T).
\end{equation*}
As  $f_i(\T_i)$ is independent to $\T_j$ for $j \neq i$, the derivative can be written in a block diagonal form
\begin{equation}\label{eqt:gra1}
\cD_{\T^*}L(\T,\lambda) =\diag\left(
 \cD_{\T_1^*}f_1(\T_1), \ldots, \cD_{\T_K^*}f_K(\T_K) \right) -\lambda \cD_{\T^*} g(\T).
\end{equation}
 
The gradient of the objective function $f_i(\T_i)$ with respect to $\T_i^*$ is
\begin{equation}
 \cD_{\T_i^*} f_i(\T_i)=\cD_{\T_i^*}\cC \left( \X_i + \bar{\T}_i \Z_i \bar{\T}_i^H \right)
-\cD_{\T_i^*}\cC \left( \X_i + \bar{\T}_i \Y_i \bar{\T}_i^{\her} \right).
\end{equation}
We begin with 
\begin{equation}\label{eqt:gradient1}
 \begin{aligned}
 & \ln(2) \cD_{\T_i^*}\cC \left( \X_i + \bar{\T}_i \Z_i \bar{\T}_i^H \right)\\ 
&= \cD_{\bar{\T}_i^*}\ln \det\left( \X_i + \bar{\T}_i \Z_i \bar{\T}_i^H \right)\cdot\cD_{\T_i^*}\bar{\T}_i^*\\
&= \bvec\left(\left( \X_i + \bar{\T}_i \Z_i \bar{\T}_i^H \right)^{-1} \bar{\T}_i \Z_i \right)^{\tran}\cdot \frac{\partial \bvec(\bar{\T}_i^*)}{\partial \bvec(\T_i^*)}\\
&= \bvec\left(\left( \X_i + \bar{\T}_i \Z_i \bar{\T}_i^H \right)^{-1} \bar{\T}_i \Z_i \right)^{\tran}\left[ \begin{array}{c}
                                                                                   \I_{M^2}\\
\0_{M^2}
                                                                                  \end{array}
\right]\\
&= \left[\left( \X_i + \bar{\T}_i \Z_i \bar{\T}_i^H \right)^{-1} \bar{\T}_i \Z_i \right]_{(:,1: M)}\\
&= \left( \X_i + \bar{\T}_i \Z_i \bar{\T}_i^H \right)^{-1} \bar{\T}_i \left[\begin{array}{c}
                                                                       \I_{M} +\tilde{\F}_{ii}\\
-\sum_{m =1}^K\bH_{im} \bP_m \tilde{\F}_{mi}
                                                                      \end{array}\right].
 \end{aligned}
\end{equation}
Similarly, we have
\begin{equation}\label{eqt:gradient2}
\begin{aligned}
& \ln(2) \cD_{\T_i^*}\cC \left( \X_i + \bar{\T}_i \Y_i \bar{\T}_i^H \right)\\
&=\left( \X_i + \bar{\T}_i \Y_i \bar{\T}_i^H \right)^{-1} \bar{\T}_i \left[\begin{array}{c}
                                                                       \tilde{\F}_{ii}\\
-\sum_{m=1}^K \bH_{im} \bP_m \tilde{\F}_{mi}
                                                                      \end{array}\right].
\end{aligned}
\end{equation}
Thus, we have the gradient of $f_i(\T_i)$ as
\begin{equation}\label{eqt:gra2}
\begin{aligned}
 & \cD_{\T_i^*} f_i(\T_i)\\
&=\frac{1}{\ln(2)}\left(\left( \X_i + \bar{\T}_i \Z_i \bar{\T}_i^H \right)^{-1} \bar{\T}_i \left[\begin{array}{c}
                                                                       \I_{M} +\tilde{\F}_{ii}\\
-\sum_{m =1}^K\bH_{im} \bP_m \tilde{\F}_{mi}
                                                                      \end{array}\right] \right.\\
& - \left(\left( \X_i + \bar{\T}_i \Y_i \bar{\T}_i^H \right)^{-1} \bar{\T}_i \left[\begin{array}{c}
                                                                       \tilde{\F}_{ii}\\
-\sum_{m=1}^K \bH_{im} \bP_m \tilde{\F}_{mi}
                                                                      \end{array}\right]\right).
\end{aligned}
\end{equation}
The last step of computing the gradient of the Lagrangian is to compute 
\begin{equation}\label{eqt:gradient3}
\begin{aligned}
 &\cD_{\T^*}\tr \left( \G^{\her \dagger} \left( \T- \bH \bP\right) \left( \tilde{\F}+ \I_{KM}\right) \left( \T- \bH \bP\right)^{\her}
\G^{ \dagger}\right)\\
&= \G^{ \dagger}\G^{\her \dagger} \left( \T- \bH \bP\right) 
\left( \tilde{\F}+ \I_{KM}\right).
\end{aligned}
\end{equation}
Combining \eqref{eqt:gra1}, \eqref{eqt:gra2} and \eqref{eqt:gradient3}, the gradient of the Lagrangian is obtained.

\bibliographystyle{IEEEbib}
\bibliography{bib2} 

\end{document}